\documentclass[12pt]{article}

\makeatletter

\def\paragraph{\@startsection{paragraph}{4}{\z@}{+2.00ex plus
 +1ex minus +.2ex}{1.5ex plus .2ex}{\it\normalsize}}

\def\section{\@startsection {section}{1}{\z@}{+3.0ex plus +1ex minus
  +.2ex}{2.3ex plus .2ex}{\normalsize\bf\boldmath}}
\def\subsection{\@startsection{subsection}{2}{\z@}{+2.5ex plus +1ex
minus +.2ex}{1.5ex plus .2ex}{\normalsize\bf\boldmath}}
\def\subsubsection{\@startsection{subsubsection}{3}{\z@}{+3.25ex plus
 +1ex minus +.2ex}{1.5ex plus .2ex}{\normalsize\bf\boldmath}}

\expandafter\ifx\csname mathrm\endcsname\relax\def\mathrm#1{{\rm #1}}\fi

\newcounter{saveeqn}

\@addtoreset{equation}{section}

\usepackage{graphicx,xspace}

\newcommand{\beq}{\begin{equation}}
\newcommand{\eeq}{\end{equation}}
\newcommand{\bea}{\begin{eqnarray}}
\newcommand{\eea}{\end{eqnarray}}
\newcommand{\bseq}{\begin{subequations}}
\newcommand{\eseq}{\end{subequations}}
\newcommand{\bsea}{\begin{subeqnarray}}
\newcommand{\esea}{\end{subeqnarray}}
\newcommand{\bit}{\begin{itemize}}
\newcommand{\eit}{\end{itemize}}
\newcommand{\ben}{\begin{enumerate}}
\newcommand{\een}{\end{enumerate}}
\newcommand{\bfig}{\begin{figure}}
\newcommand{\efig}{\end{figure}}
\newcommand{\btab}{\begin{table}}
\newcommand{\etab}{\end{table}}
\newcommand{\im}[1]{\ensuremath{#1}\xspace}       
\newcommand{\imx}[1]{\ensuremath{#1}\xspace}       
\newcommand{\etal}{{\em et al.}}

\newcommand{\ie}{{\em i.e.}}

\newcommand{\eV}{\imx{\mathrm{e\kern -0.1em V}}}
\newcommand{\MeV}{\imx{\mathrm{Me\kern -0.1em V}}}
\newcommand{\GeV}{\imx{\mathrm{Ge\kern -0.1em V}}}
\newcommand{\TeV}{\imx{\mathrm{Te\kern -0.1em V}}}

\newcommand{\IM}{\im{\mathrm{I} \kern -0.15 em \mathrm{m}}}         
\newcommand{\RE}{\im{\mathrm{R} \kern -0.15 em \mathrm{e}}}         

\newcommand{\Pchi}{\im{{\raise5pt\hbox{$\chi$}}}}
\newcommand{\Pe}{\im{\mathrm{e}}}            
\newcommand{\Pm}{\im{\mu}}
\newcommand{\Pt}{\im{\tau}}
\newcommand{\Pn}{\im{\nu}}
\newcommand{\Pl}{\im{\ell}}
\newcommand{\Pp}{\im{\mathrm{p}}}
\newcommand{\Pu}{\im{\mathrm{u}}}
\newcommand{\Pd}{\im{\mathrm{d}}}

\newcommand{\Pc}{\im{\mathrm{c}}}
\newcommand{\Pb}{\im{\mathrm{b}}}
\newcommand{\PT}{\im{\mathrm{t}}}
\newcommand{\Pq}{\im{q}}
\newcommand{\Pf}{\im{f}}
\newcommand{\Phad}{\im{\mathrm{had}}}

\newcommand{\Plep}{\im{\mathrm{lep}}}

\newcommand{\PW}{\im{\mathrm{W}}}             

\newcommand{\PZ}{\im{\mathrm{Z}}}
\newcommand{\PH}{\im{\mathrm{H}}}

\newcommand{\MW}{\im{M_{\PW}}}
\newcommand{\MZ}{\im{M_{\PZ}}}
\newcommand{\MH}{\im{M_{\PH}}}

\newcommand{\MT}{\im{M_{\PT}}}
\newcommand{\Nn}{\im{N_{\Pn}}}
\newcommand{\G}{\im{\Gamma}}                  
\newcommand{\GZ}{\im{\G_{\PZ}}}               

\newcommand{\GZb}{\im{\G_{\Pb\Pb}}}
\newcommand{\GZq}{\im{\G_{\Pq\Pq}}}

\newcommand{\GZhad}{\im{\G_{\Phad}}}

\newcommand{\GW}{\im{\G_{\PW}}}               

\newcommand{\A}{\im{\mathrm{A}}}
\newcommand{\V}{\im{\mathrm{V}}}
\newcommand{\R}{\im{\mathrm{R}}}

\newcommand{\g}{\im{g}}                       

\newcommand{\gae}{\im{\g_{\A\Pe}}}

\newcommand{\gvq}{\im{\g_{\V\Pq}}}

\newcommand{\gaf}{\im{\g_{\A\Pf}}}
\newcommand{\gvf}{\im{\g_{\V\Pf}}}

\newcommand{\Ae}{\im{\A_{\Pe}}}

\newcommand{\Al}{\im{\A_{\Pl}}}
\newcommand{\Ab}{\im{\A_{\Pb}}}

\newcommand{\Aq}{\im{\A_{\Pq}}}
\newcommand{\Af}{\im{\A_{\Pf}}}

\newcommand{\Afb}{\im{\A_{\mathrm{fb}}}}

\newcommand{\Afbzl}{\im{\Afb^{0,\Pl}}}
\newcommand{\Afbzb}{\im{\Afb^{0,\Pb}}}
\newcommand{\Afbzc}{\im{\Afb^{0,\Pc}}}

\newcommand{\RZl}{\im{\R^{\PZ}_{\Pl}}}
\newcommand{\RZb}{\im{\R^{\PZ}_{\Pb}}}

\newcommand{\Rb}{\im{\R_{\Pb}}}

\newcommand{\Rq}{\im{\R_{\Pq}}}

\newcommand{\swsqeffl}{\sin^2\theta_{\mathrm{eff}}^{\mathrm{lept}}}

\newcommand{\Pee}{\im{\Pe^+\Pe^-}}
\newcommand{\Pmm}{\im{\Pm^+\Pm^-}}
\newcommand{\Ptt}{\im{\Pt^+\Pt^-}}
\newcommand{\Pll}{\im{\Pl^+\Pl^-}}
\newcommand{\Ppp}{\im{\Pp\overline{\Pp}}}

\newcommand{\Pnn}{\im{\Pn\overline{\Pn}}}
\newcommand{\Pqq}{\im{\Pq\overline{\Pq}}}
\newcommand{\PTT}{\im{\PT\overline{\PT}}}
\newcommand{\Pbb}{\im{\Pb\overline{\Pb}}}

\newcommand{\Pff}{\im{\Pf\overline{\Pf}}}
\newcommand{\PWW}{\im{\PW^+\PW^-}}
\newcommand{\PZZ}{\im{\PZ\PZ}}

\newcommand{\Peeee}{\im{\Pee \kern -0.35em \rightarrow\Pee}}
\newcommand{\Peemm}{\im{\Pee \kern -0.35em \rightarrow\Pmm}}
\newcommand{\Peett}{\im{\Pee \kern -0.35em \rightarrow\Ptt}}
\newcommand{\Peell}{\im{\Pee \kern -0.35em \rightarrow\Pll}}
\newcommand{\Peenn}{\im{\Pee \kern -0.35em \rightarrow\Pnn}}
\newcommand{\Peeqq}{\im{\Pee \kern -0.35em \rightarrow\Pqq}}
\newcommand{\Peehad}{\im{\Pee \kern -0.35em \rightarrow\Phad}}
\newcommand{\Peeff}{\im{\Pee \kern -0.35em \rightarrow\Pff}}
\newcommand{\PeeTT}{\im{\Pee \kern -0.35em \rightarrow\PTT}}
\newcommand{\PeeWW}{\im{\Pee \kern -0.35em \rightarrow\PWW}}
\newcommand{\PeeZZ}{\im{\Pee \kern -0.35em \rightarrow\PZZ}}
\newcommand{\Pffff}{\im{\Pff\Pff}}
\newcommand{\Pqqqq}{\im{\Pqq\Pqq}}

\newcommand{\aqcd}{\im{\alpha_S}}

\newcommand{\GF}{\im{G_{\mathrm{F}}}}

\newcommand{\dalhad}{\im{\Delta\alpha^{(5)}_{had}}}
\newcommand{\RH}{\im{R_\PH}}
\newcommand{\QW}{\im{Q_{\mathrm{W}}}}

\begin{document}

\begin{titlepage} 
\noindent                
{\Large 15 April 2003    \hfill UCD-EXPH/030401\\
 $\phantom{0}$           \hfill  hep-ex/0304023\\}
\begin{center}
\vskip -1cm
\includegraphics[width=4cm]{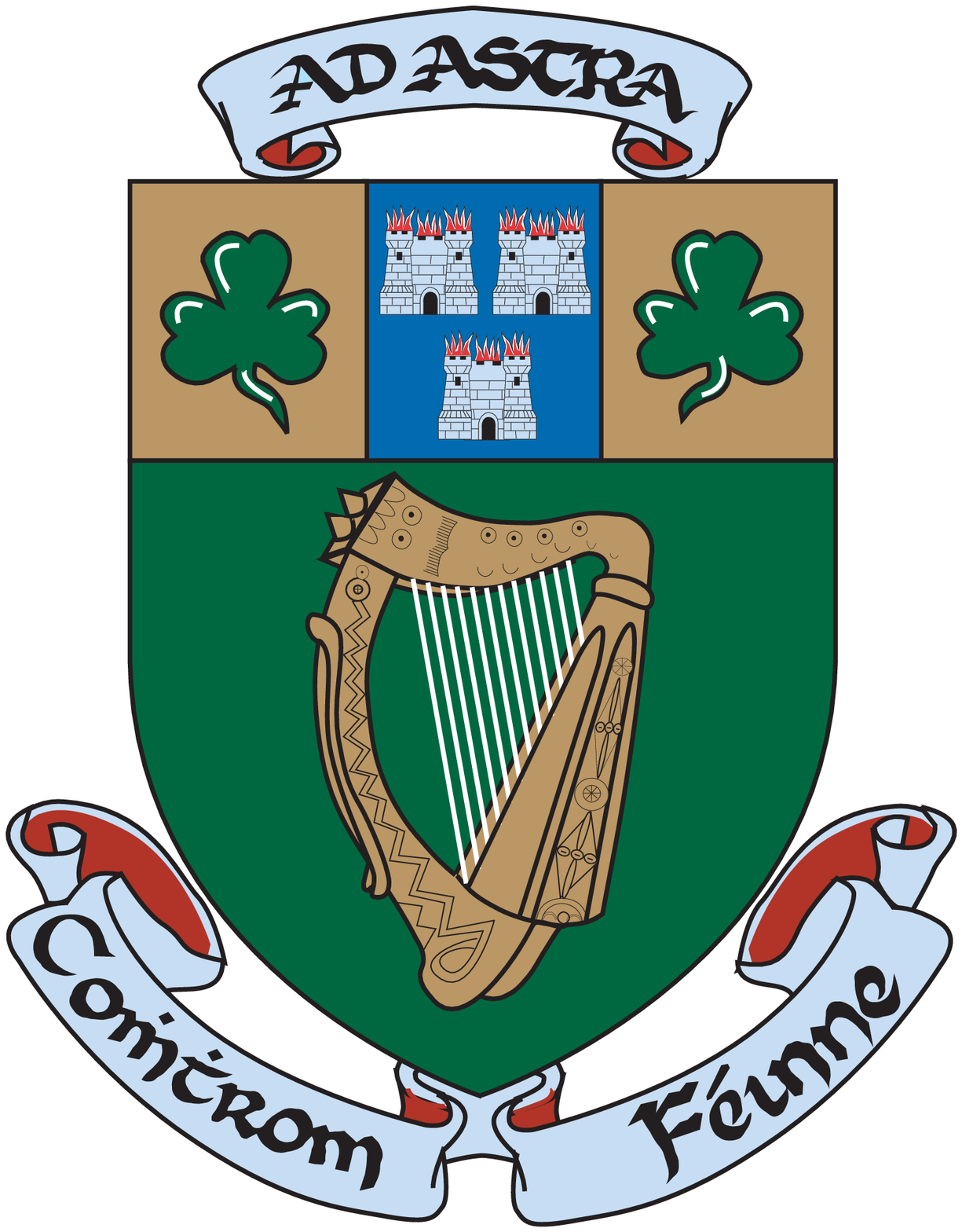}
\vskip 1cm            
{\Huge\bf Electroweak Precision Data \\
           Global Higgs Analysis \\}
\vskip 1cm
{\Large {\bf Martin W. Gr\"unewald}\\[20pt]
        Department of Experimental Physics\\
        University College Dublin\\
        Belfield, Dublin 4\\
        Ireland\\}
\vfill
{\bf Abstract} \\[10pt]
\end{center}
{
  
  The status of published and preliminary precision electroweak
  measurements as of winter 2002/03 is presented. The new results on
  the mass of the W boson as measured at LEP-2 and on atomic parity
  violation in Caesium are included.  The experimental results are
  compared with the predictions of the minimal Standard Model and are
  used to constrain its parameters, including the mass of the Higgs
  boson. The agreement between measurements and expectations from
  theory is discussed.

}
\vskip 1cm
\begin{center}
\em{ Invited talk presented at the Mini-Workshop\\
 ``ELECTROWEAK PRECISION DATA AND THE HIGGS MASS'' \\
  DESY Zeuthen, Germany, February 28th to March 1st, 2003}
\end{center}

\end{titlepage}
\clearpage
\begin{titlepage}
$ $
\end{titlepage}
\clearpage

\setcounter{page}{1}

\vspace*{1cm}
\begin{center}

{\Large \bf   Electroweak Precision Data - Global Higgs Analysis }

\vspace*{1cm}

{\sc Martin W. Gr\"unewald}

\vspace*{.5cm}

{\normalsize \it
        Department of Experimental Physics,
        University College Dublin,\\
        Belfield, Dublin 4,
        Ireland
}
\par
\end{center}
\vskip 1cm
\begin{center}
\bf Abstract
\end{center} 
{\it

  The status of published and preliminary precision electroweak
  measurements as of winter 2002/03 is presented. The new results on
  the mass of the W boson as measured at LEP-2 and on atomic parity
  violation in Caesium are included.  The experimental results are
  compared with the predictions of the minimal Standard Model and are
  used to constrain its parameters, including the mass of the Higgs
  boson. The agreement between measurements and expectations from
  theory is discussed.

}
\par
\vskip 1cm

\section{Introduction}

On the level of realistic observables such as measured cross sections,
ratios and asymmetries, the set of electroweak precision data, all
obtained within the last 15 years, consists of over thousand
measurements with partially correlated statistical and systematic
uncertainties. This large set of results is reduced to a more
manageable set of twenty precision so-called pseudo observables in a
largely model-independent procedure, allowing for tests of the
Standard Model (SM) and other theories of nature at the fundamental
level.

About 2/3 of all pseudo observables arise from measurements performed
in electron-positron collisions at the Z pole, by the SLC experiment
SLD and the four LEP experiments ALEPH, DELPHI, L3 and OPAL.  The
Z-pole observables are: 5 observables describing the Z lineshape and
leptonic forward-backward asymmetries, 2 observables describing
polarised leptonic asymmetries measured by SLD with polarised beams
and at LEP exploiting tau polarisation, 6 observables describing
b-quark and c-quark pair production at the Z pole, and the inclusive
hadronic charge asymmetry.  The six remaining measurements are: the
mass and total width of the W boson measured by the TEVATRON
experiments CDF and D\O\ and the four LEP-2 experiments, the top quark
mass measured at the TEVATRON, the neutrino-nucleon scattering cross
section ratio as measured by NuTeV, atomic parity violation in
Caesium, and the hadronic vacuum polarisation at the Z pole.  Also,
``constants'' such as the Fermi constant $\GF$ are used.

In the following these pseudo observables are discussed and used to
perform various Standard Model analyses~\cite{LEPEWWG}.  For the
hadronic vacuum polarisation, $\dalhad$, the reader is referred to the
dedicated workshop presentation~\cite{FredJ}.

\section{Atomic Parity Violation}

The interaction between an electron and the atomic nucleus receives a
parity-violating contribution due to $\gamma$/Z interference. The weak
charge, $\QW$, of the nucleus is measured:
\begin{eqnarray}
\QW(Z,N) & = & -2\left[\left(2Z+N\right)C_{1u} +
                       \left(Z+2N\right)C_{1d}\right]\,,
\end{eqnarray}
with $C_{1q}=2\gae\gvq$ for $q=\Pu,\Pd$ in the limit of zero momentum
transfer.  Thus $\QW$ is defined simply as the sum of the weak charges
of up- and down-quark in a nucleus containing $Z$ protons and $N$
neutrons and thus $(2Z+N)$ up-quarks and $(Z+2N)$ down-quarks.
Therefore, the raw measurement must be corrected for nuclear many-body
effects and QED radiative corrections.  The most precise measurement
is performed for Caesium~\cite{APV-Caesium}.  Recent progress in QED
self-energy and vertex radiative corrections to order $Z\alpha^2$ and
$Z^2\alpha^3$ results in a significant shift in the experimental
result for $\QW$; the new result~\cite{APV-Caesium}:
\begin{eqnarray}
\QW(55,78) & = & -72.83\pm0.29~(exp.)\pm0.39~(theo.)\,,
\end{eqnarray}
is now in perfect agreement with the SM expectation.

\section{Neutrino Nucleon Scattering}

The NuTeV collaboration studies $t$-channel neutrino-nucleon
scattering at an average momentum transfer of $20~\GeV$, analysing
both charged current (CC) and neutral current (NC) reactions.  Using
both a neutrino and an anti-neutrino beam with high statistics, it is
possible to exploit the Paschos-Wolfenstein relation~\cite{PWR}:
\begin{eqnarray}
R_-  =  \frac{\sigma_{NC}(\nu)-\sigma_{NC}(\bar\nu)}
               {\sigma_{CC}(\nu)-\sigma_{CC}(\bar\nu)}
 =  4 g^2_{L\nu}\sum_{u,d}\left[g^2_{Lq} - g^2_{Rq}\right]
 =  \rho_\nu\rho_{ud}\left[1/2 -
               \sin^2\theta_W^{on-shell}\right]\,,
\end{eqnarray}
where the sum runs over the valence quarks, u and d. This relation
holds for iso-scalar targets and up to small electroweak radiative
corrections.  In the ideal case this measurement is insensitive to the
effects of sea quarks, which cancel. Charm production, uncertain due
to charm mass effects, enters only through CC scattering off valence d
quarks, a CKM suppressed process.  Using $\nu_\mu/\bar\nu_\mu$ beams,
CC reactions are discriminated from NC reactions by the presence of a
primary $\mu^-/\mu^+$ in the final state.

Assuming $\rho=\rho_{SM}$, NuTeV's final results reads~\cite{NuTeV}:
\begin{eqnarray}
\sin^2\theta_W^{on-shell} & \equiv & 1-\MW^2/\MZ^2 
~ = ~ 0.2277 \pm 0.0013  \pm 0.0009  \\
&   & - 0.00022 \frac{\MT^2-(175~\GeV)^2}{(50~\GeV)^2} 
      + 0.00032 \ln(\MH/150~\GeV)\,, \nonumber
\end{eqnarray}
where the first error is statistical and the second is systematic.
NuTeV's final result is still statistics limited.  The result is in
excellent agreement with the previous world average~\cite{nNWorld} but
differs by 2.9 standard deviations from the prediction of the global
SM analysis presented in Section~\ref{sec:MSM}.  Further details on
NuTeV's measurement are given in~\cite{KevinMcF}.

\begin{figure}[tbp]
\begin{center}
\includegraphics[width=0.49\linewidth]{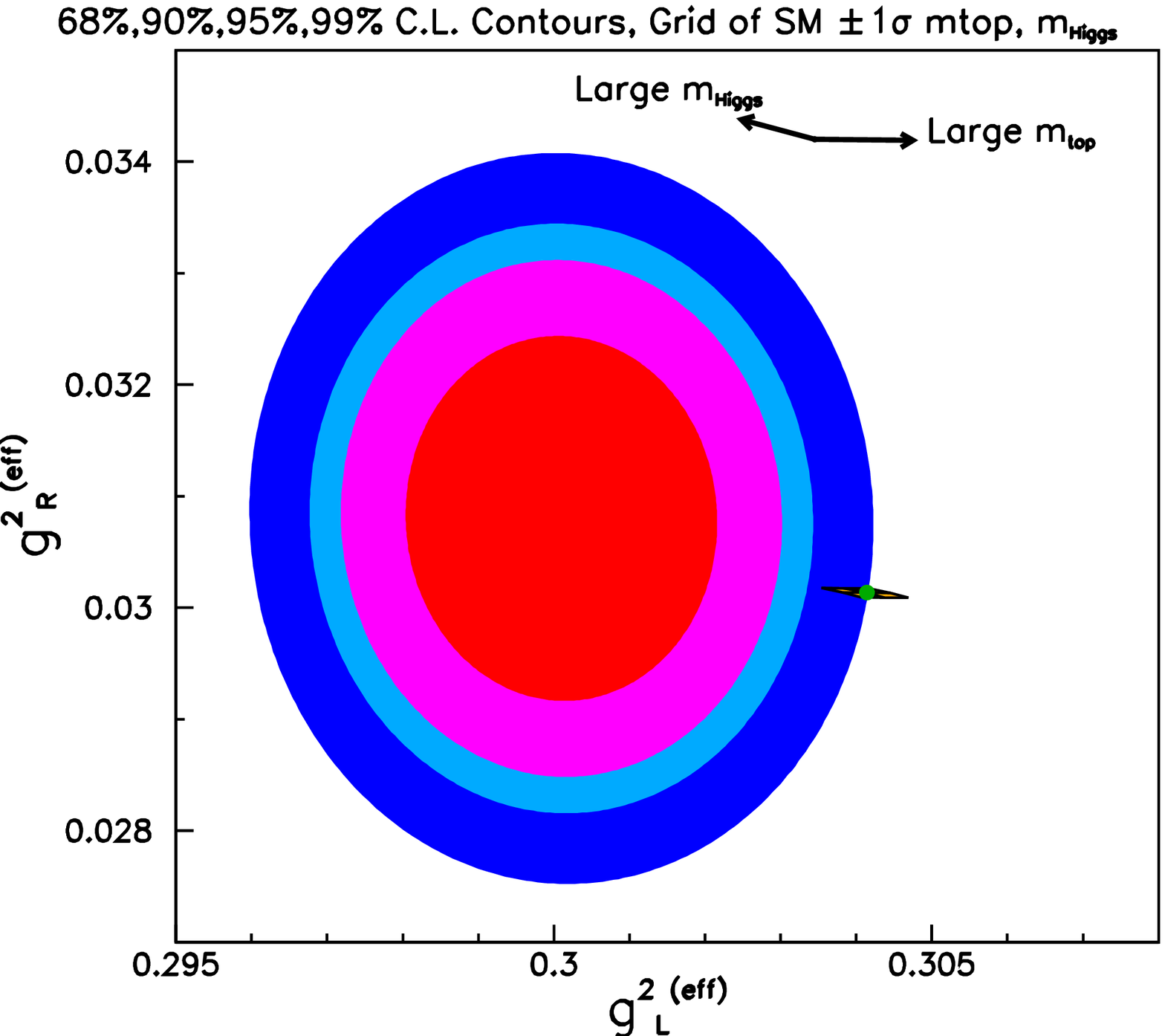}
\hfill
\includegraphics[width=0.49\linewidth]{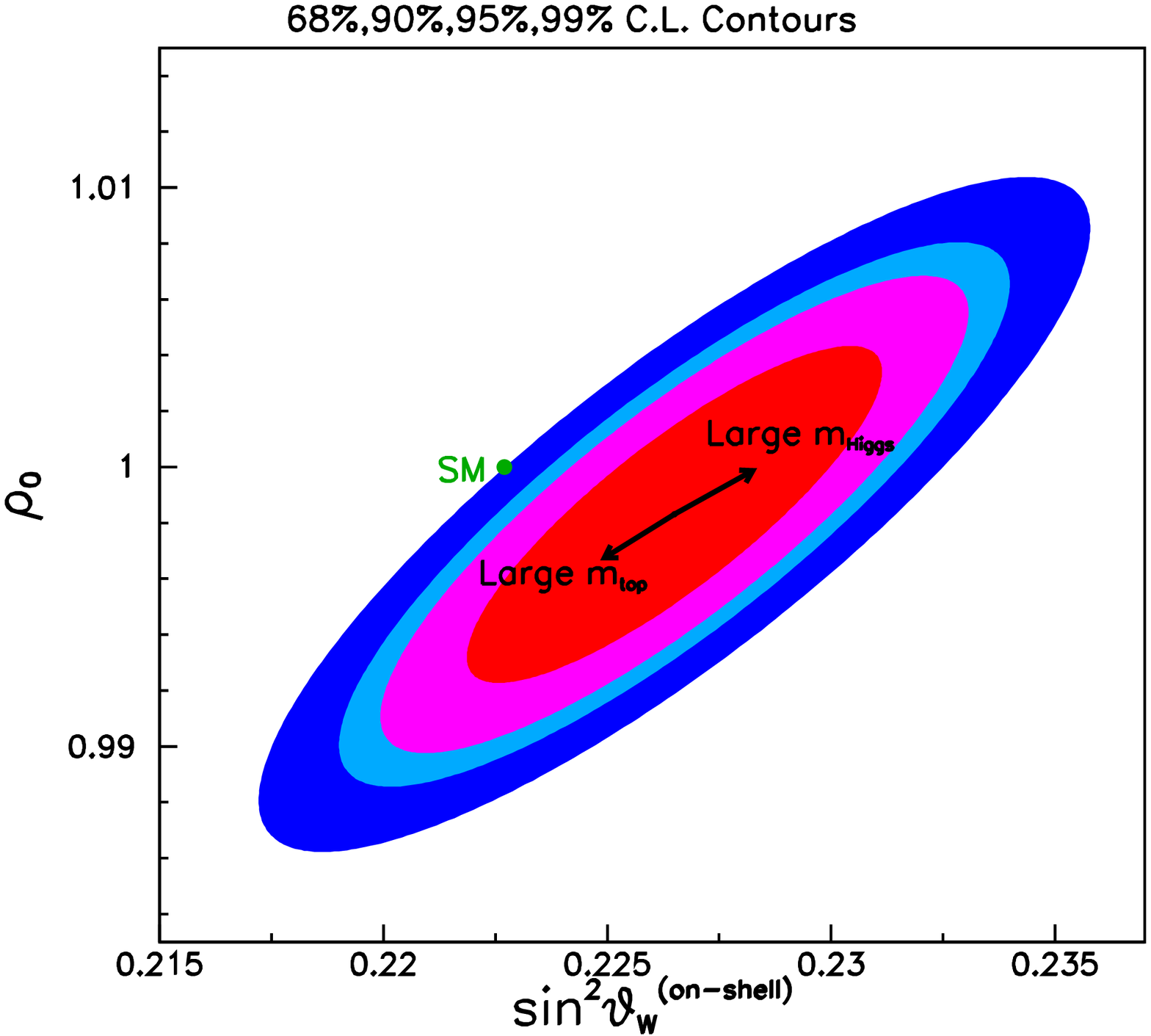}
\caption{NuTeV result in the plane of effective right- and left-handed
  couplings (left) and of on-shell angle versus $\rho$-scale factor
  $\rho_0$ (right). }
\label{fig:nutev-2d}
\end{center}
\end{figure}

When quoting the result in terms of $\sin^2\theta_W^{on-shell}$, as
done historically, it must be assumed that $\rho=\rho_{SM}$.  In a
more flexible ansatz allowing model-independent interpretations, the
NuTeV result is also presented in terms of effective left- and
right-handed couplings, shown in Figure~\ref{fig:nutev-2d} (left) and
defined as: $g^{2~(\mathrm{eff})}_X = 4g^2_{L\nu}\sum_q g^2_{Xq} $ for
$X=L,R$.  Here the deviation is confined to the effective left-handed
coupling product.  Modifying all $\rho$ parameters by a scale factor
$\rho_0$ shows that either $\rho_0$ or the mixing angle, but not both,
could be in agreement with the SM, as visible in
Figure~\ref{fig:nutev-2d} (right).  Assuming the electroweak mixing
angle to have it's expected value, the change in the $\rho$ factors
can be absorbed in $\rho_\nu$, \ie, interpreted as a change in the
coupling strength of neutrinos, then lower than expected by about
$(1.2\pm0.4)\%$.  A similar trend is observed with the neutrino
coupling as measured by the invisible width of the Z boson at LEP-1,
yielding a smaller and less significant deficit of $(0.5\pm0.3)\%$ in
$\rho_\nu$.  Various explanations besides being a statistical
fluctuation, have been put forward and reviewed at this workshop,
ranging from old and new physics effects, but it seems none is able to
explain the result naturally~\cite{KevinMcF,SMoch}.

\section{Mass of the Top Quark}

\begin{figure}[tbp]
\begin{center}
\includegraphics[width=0.49\linewidth]{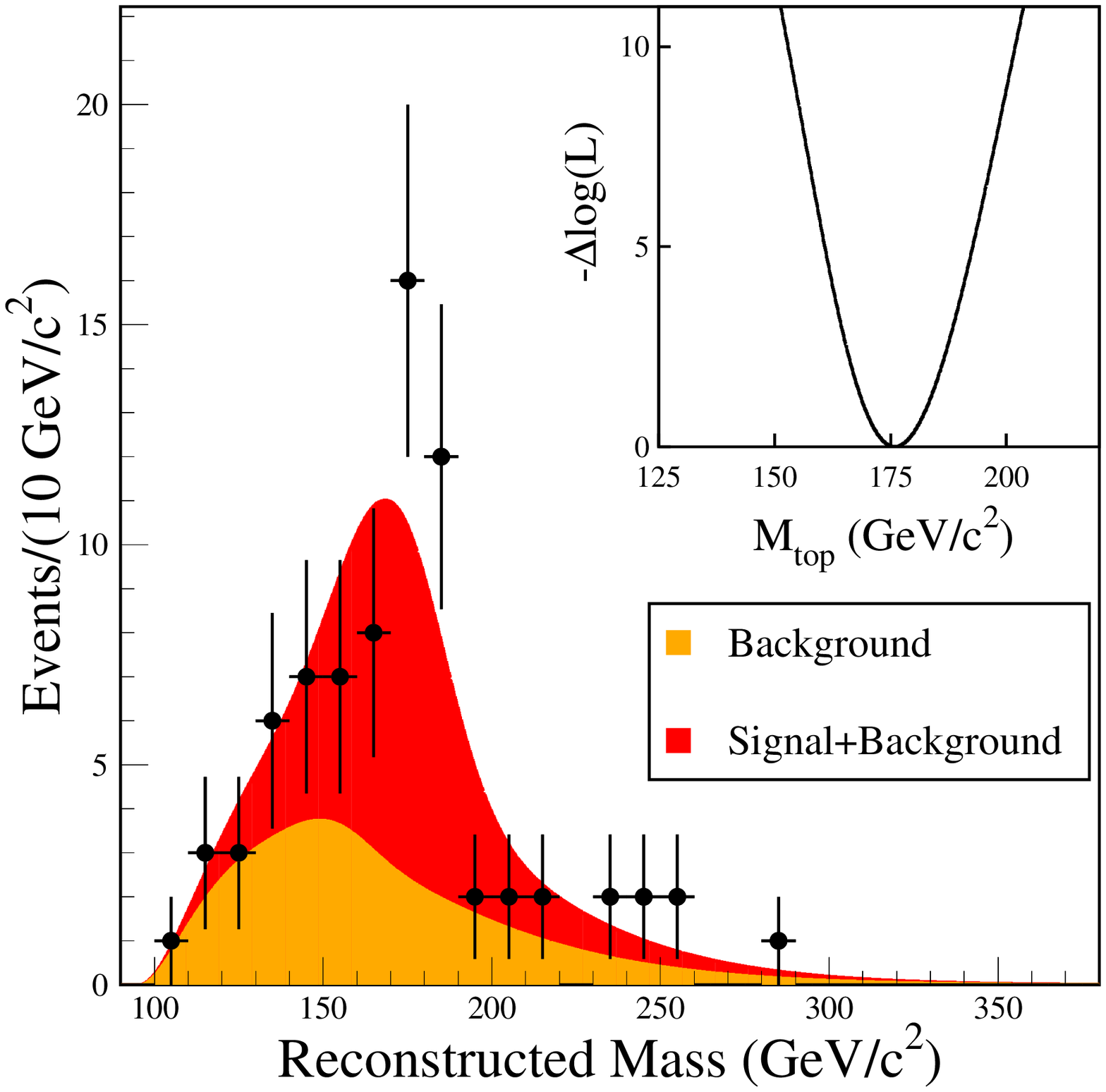}
\hfill
\includegraphics[width=0.49\linewidth]{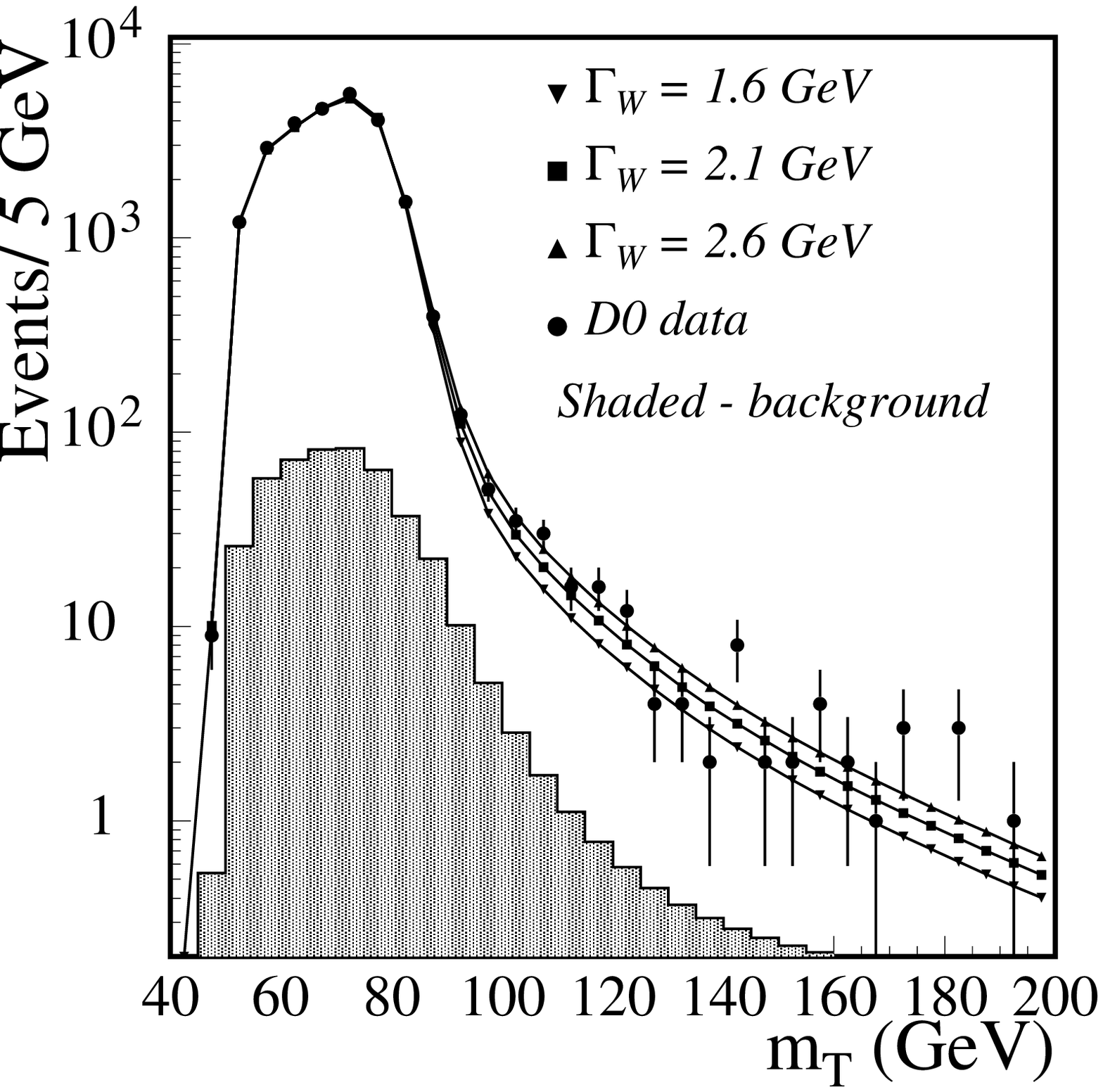}
\caption{Left: Reconstructed top-quark mass in $\PTT$ events as
measured by CDF. Right: Reconstructed transverse mass in $\PW$ events
as measured by D\O. }
\label{fig:tev-mt-mw}
\end{center}
\end{figure}

In 1995, the TEVATRON experiments CDF and D\O\ discovered the top
quark in proton-antiproton collisions at $1.8~\TeV$ centre-of-mass
energy, by observing the reaction $\Ppp\to\PTT~X,~\PTT\to\Pbb\PWW$.
Depending on the decay modes of the two W bosons, the event signatures
are two b-jets plus either di-leptons, lepton plus jets, or all jets.
The distribution of the reconstructed top-quark mass as measured by
CDF is shown in Figure~\ref{fig:tev-mt-mw} (left).  The results
published based on data collected in Run-I are combined taking
correlated systematic uncertainties into account~\cite{RPP2002}:
\begin{eqnarray}
\MT & = & 174.3\pm3.2~(stat.)\pm4.0~(syst.)~\GeV\,.
\end{eqnarray}
The D\O\ collaboration has recently presented a new preliminary Run-I
based result in the lepton-plus-jets channel, which has a reduced
uncertainty and a central value a few GeV higher than that of their
previous analysis in this channel entering the above
average~\cite{D0-MT-L+J}.

The systematic uncertainties are dominated by the jet energy scale (2
to $5~\GeV$ depending on channel), which will reduce with more
data. Signal and background modelling, including hard matrix elements,
parton distribution functions and MC generators account for a smaller
part.  Top-quark mass measurements with an uncertainty of less than
$2-3~\GeV$ are expected from the ongoing TEVATRON Run-II.

\section{Mass and Width of the W boson}

Until 1996, the W boson mass and width was measured at hadron
colliders only, most recently by the experiments CDF and D\O.
Leptonic W decays with electrons and muons are selected and
reconstructed. The transverse mass, \ie, the invariant mass of the
lepton and the missing momentum vector in the plane transverse to the
beam axis is unaffected by the unknown longitudinal boost of the W
boson and bounded from above by the invariant mass of the decaying W
boson.  The distribution of the transverse mass as measured by D\O\ is
shown in Figure~\ref{fig:tev-mt-mw} (right).  The sharp upper edge of
the so called Jacobian peak yields the mass of the W boson, while the
W-boson width is derived from the high-mass tail of this distribution.
Final results on $\MW$ and $\GW$ from CDF and D\O\ are now available
for the complete Run-I data set.  They are combined taking
correlations properly into account~\cite{TEV-MW-GW}.

The uncertainties on both mass and width of the W boson are dominated
by the limited data statistics. The largest systematic uncertainty
arises from the energy measurement of the leptons, and this energy
scale uncertainty will also reduce with more data. The signal model,
parton distribution functions, gluon radiation and QED corrections in
leptonic W decays are less important. Further details on the
measurement of $\MW$ at hadron colliders are given in~\cite{UlrichB}.
W-boson mass measurements with an uncertainty of less than $25~\MeV$
are expected from the ongoing TEVATRON Run-II.

Since 1996 the W boson mass and width is also measured at LEP-2 using
$\Pee\rightarrow\PWW\rightarrow\Pffff$ events.  Four-fermion final
states are selected and the two decaying W bosons are reconstructed.
For hadronic and semileptonic W-pair events, the W-pair kinematic is
completely reconstructed so that one directly measures the invariant
masses of the decaying W bosons.  These results are preliminary.

For hadronic W-pair events, $\Pee\rightarrow\PWW\rightarrow\Pqqqq
\rightarrow hadrons$, cross talk effects may occur between the two
hadronic systems.  The four-momentum exchange causes the mass of the
decaying W bosons to be different from the measured mass of the
hadronic decay products, thus leading to potentially large systematic
effects.  Because of these large additional uncertainties compared to
the semileptonic channel, now bounded by studies based on data, the
weight of the four jet channel in the LEP average is less than
10\%. The difference in mass obtained for hadronic and semileptonic
W-pair events, calculated without FSI uncertainties, is
$(22\pm43)~\MeV$, \ie, showing no bias.  Further details on the LEP
measurements are given in~\cite{RichardH}.

\begin{figure}[hb]
\begin{center}
\vskip -0.4cm
\vfill
\includegraphics[width=0.54\linewidth]{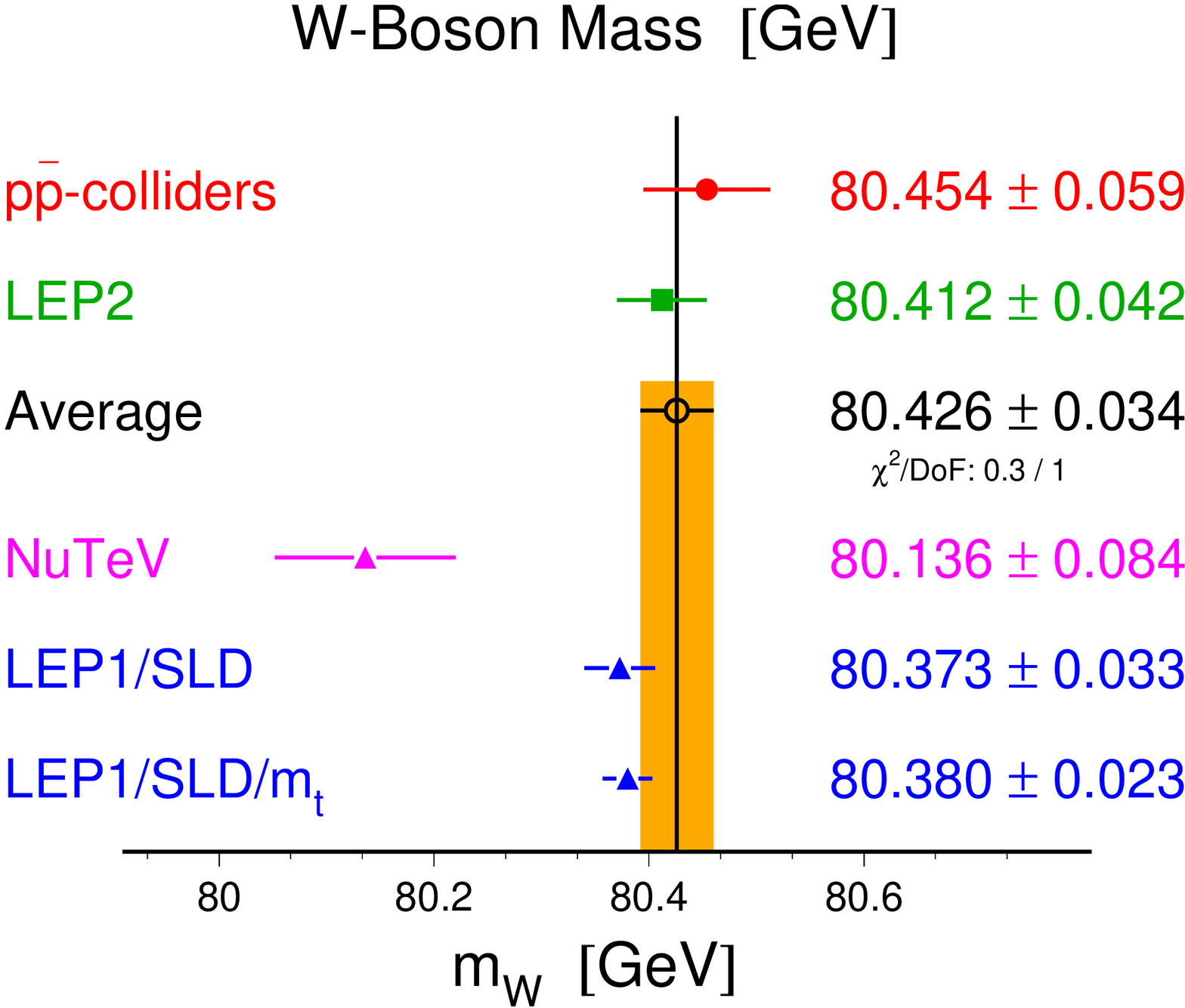}
\hfill
\includegraphics[width=0.45\linewidth]{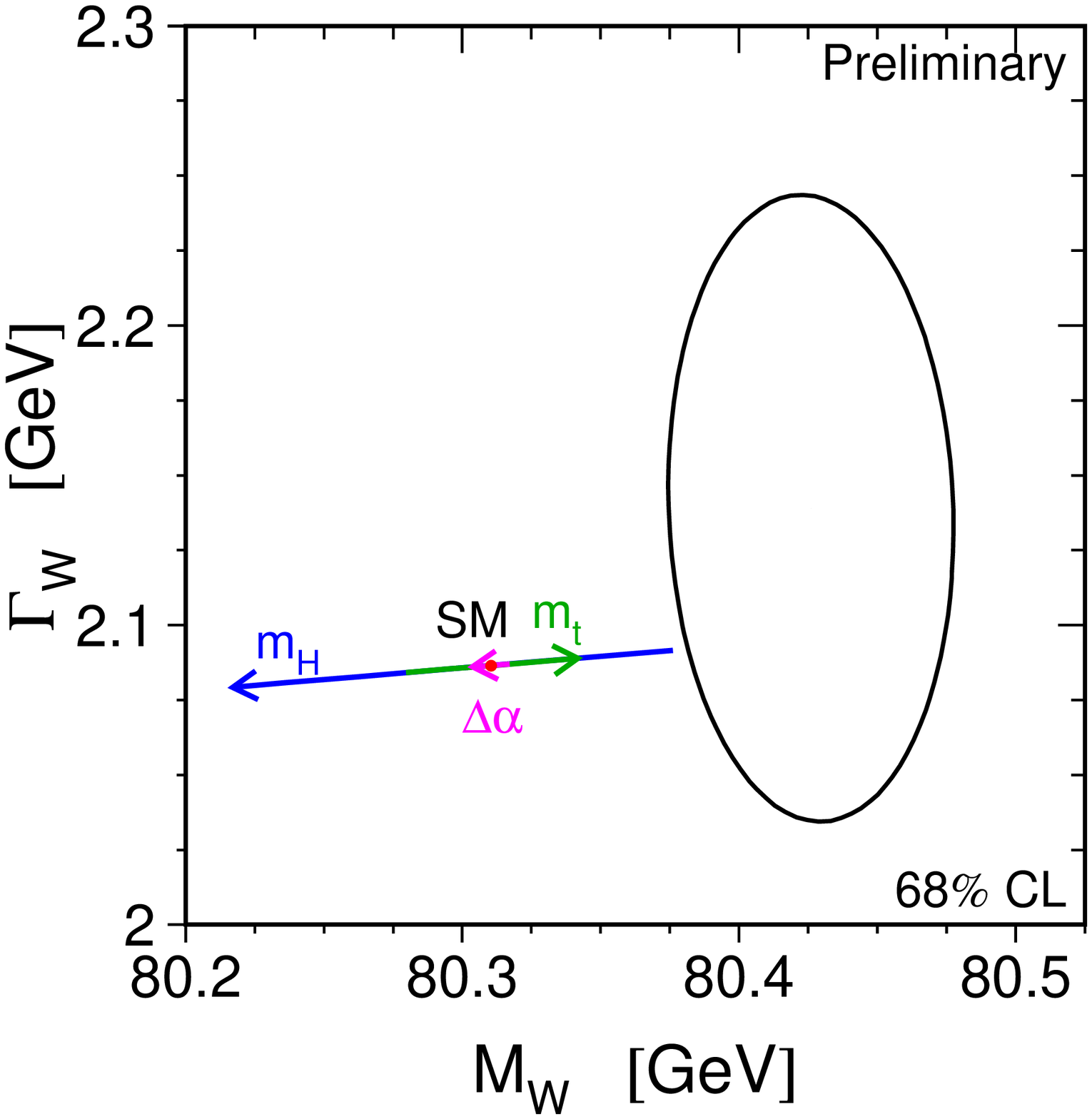}
\caption{Left: Results on $\MW$ obtained from the TEVATRON and LEP
  experiments, compared with SM expectations. Right: Contour curve of
  68\% C.L. in the $(\MW,\GW)$ plane.  The SM expectation is shown as
  the arrow for $\dalhad=0.02761\pm0.00036$, $\MT=174.3\pm5.1~\GeV$
  and $\MH=300^{+700}_{-186}~\GeV$.}
\label{fig:mw-gw-results}
\end{center}
\end{figure}

The results of the TEVATRON and LEP experiments on the mass of the W
boson are in very good agreement as shown in
Figure~\ref{fig:mw-gw-results} (left).  The combined results and their
correlation is shown in Figure~\ref{fig:mw-gw-results} (right).  It
can be seen that the W mass is highly sensitive to SM parameters, in
particular preferring a low value for the mass of the Higgs boson.

\clearpage

\section{Z-Boson Physics}

In the previous decade, electron-positron annihilations at high
energies have allowed to measure precisely a wealth of electroweak
observables related to Z-boson couplings to fermion-antifermion pairs.
These measurements are performed by the SLC experiment SLD, and the
LEP experiments ALEPH, DELPHI, L3 and OPAL.

\subsection{Z Lineshape and Leptonic F/B Asymmetries}

\begin{figure}[htbp]
\begin{center}
\includegraphics[width=0.59\linewidth]{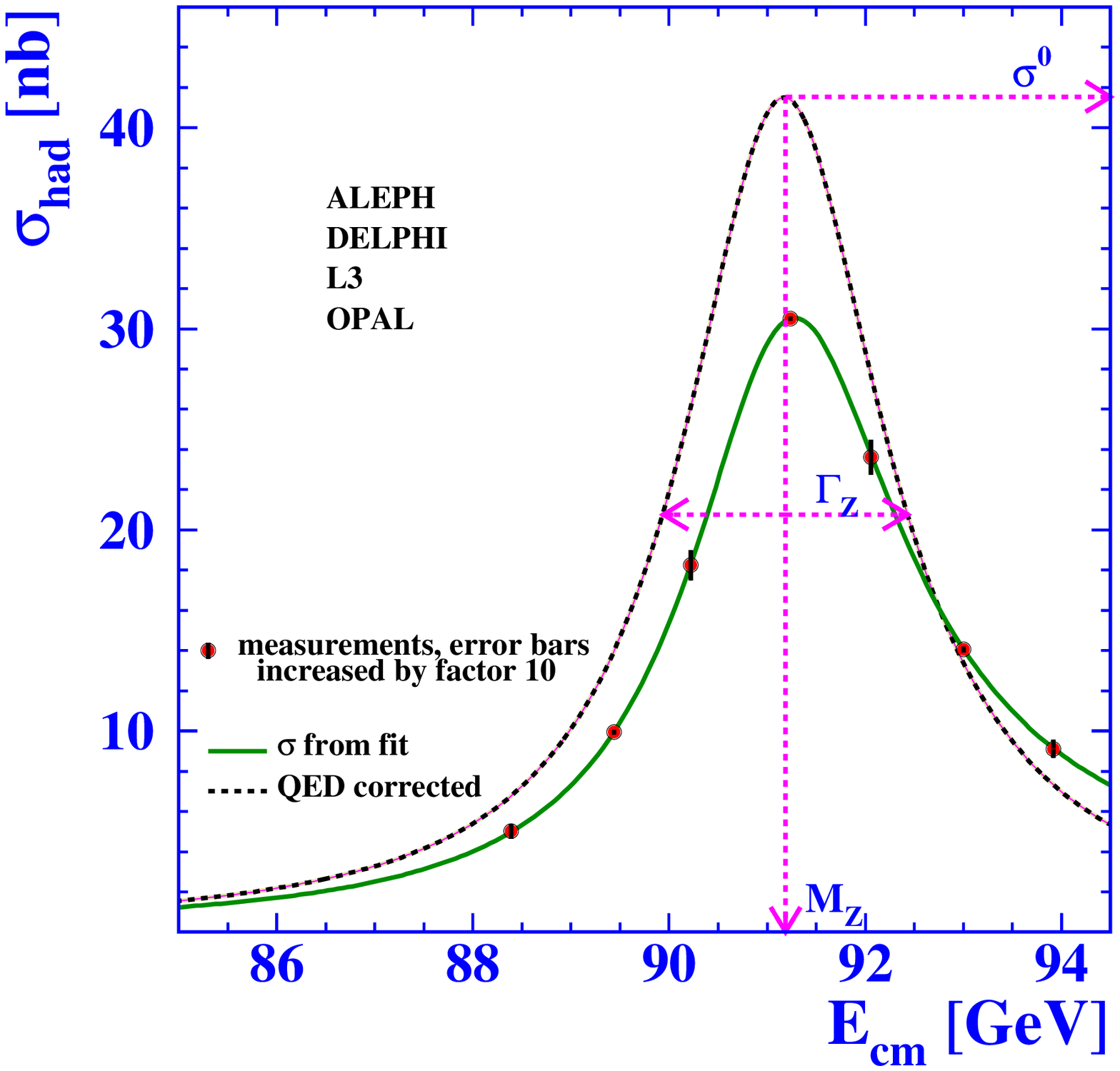}
\hfill
\includegraphics[width=0.39\linewidth]{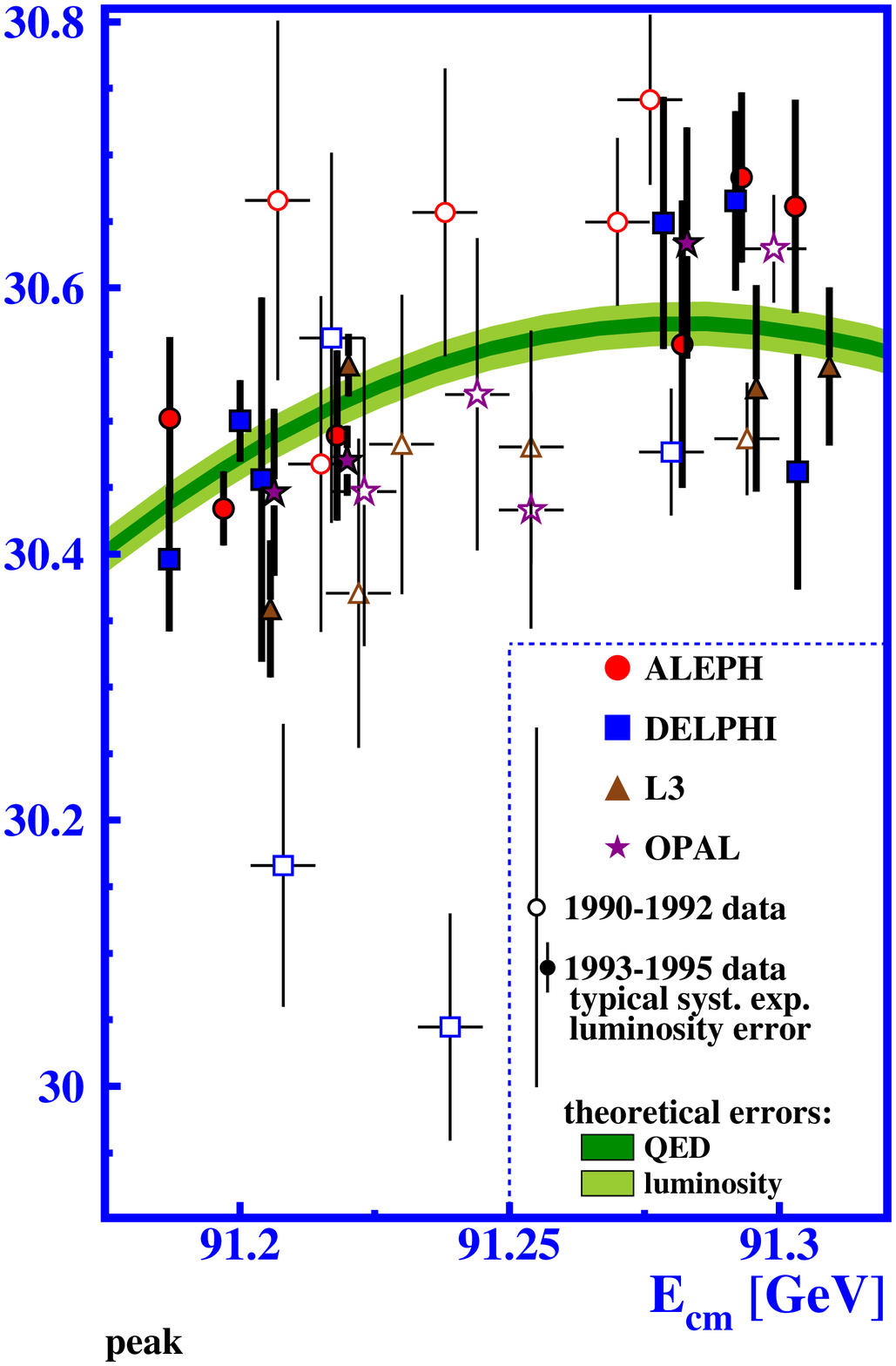}
\caption{Hadron production in $\Pee$ annihilation.}
\label{fig:ee-had}
\end{center}
\end{figure}

\begin{figure}[tbp]
\begin{center}
\includegraphics[width=0.49\linewidth]{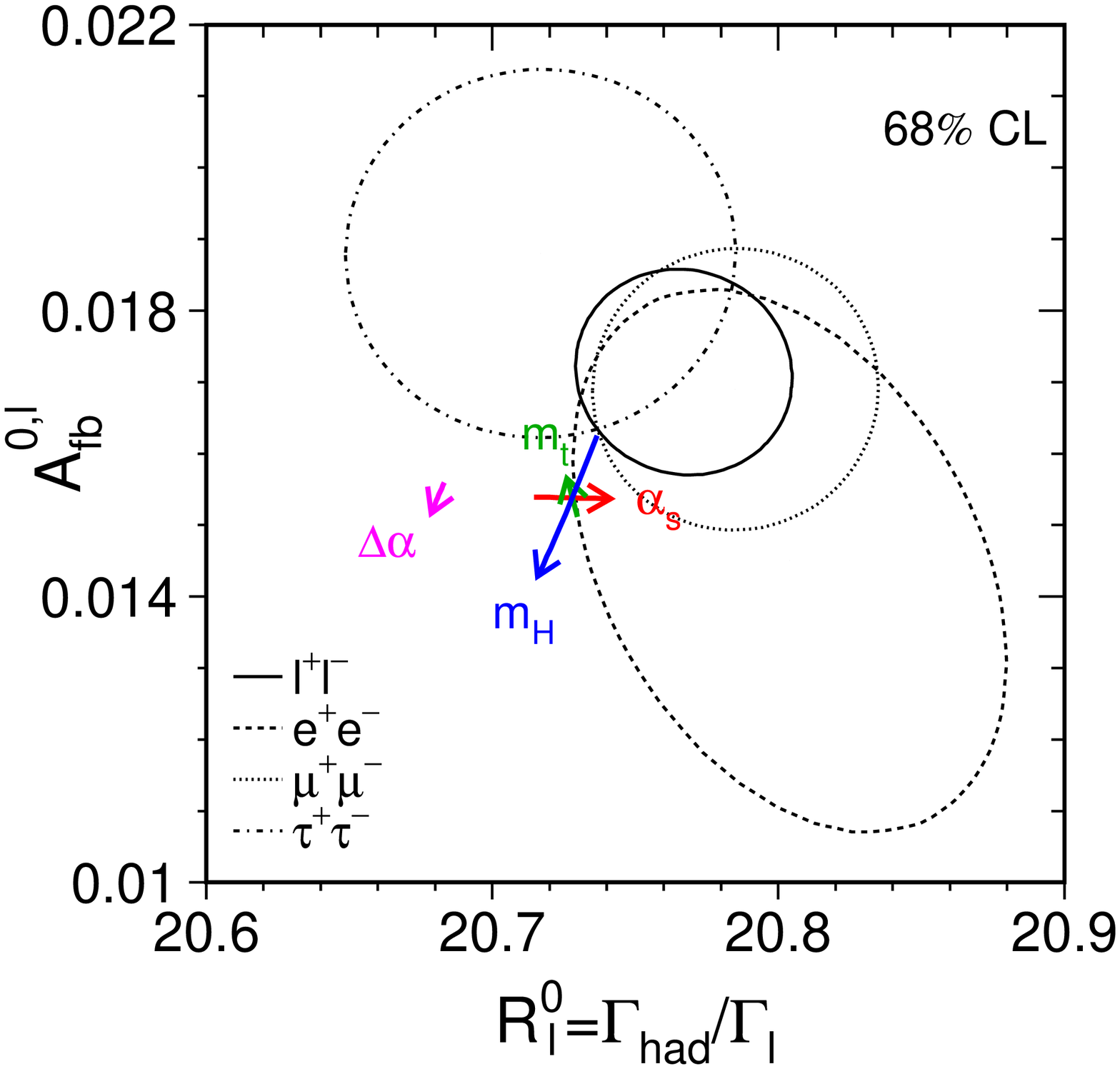}
\hfill
\includegraphics[width=0.49\linewidth]{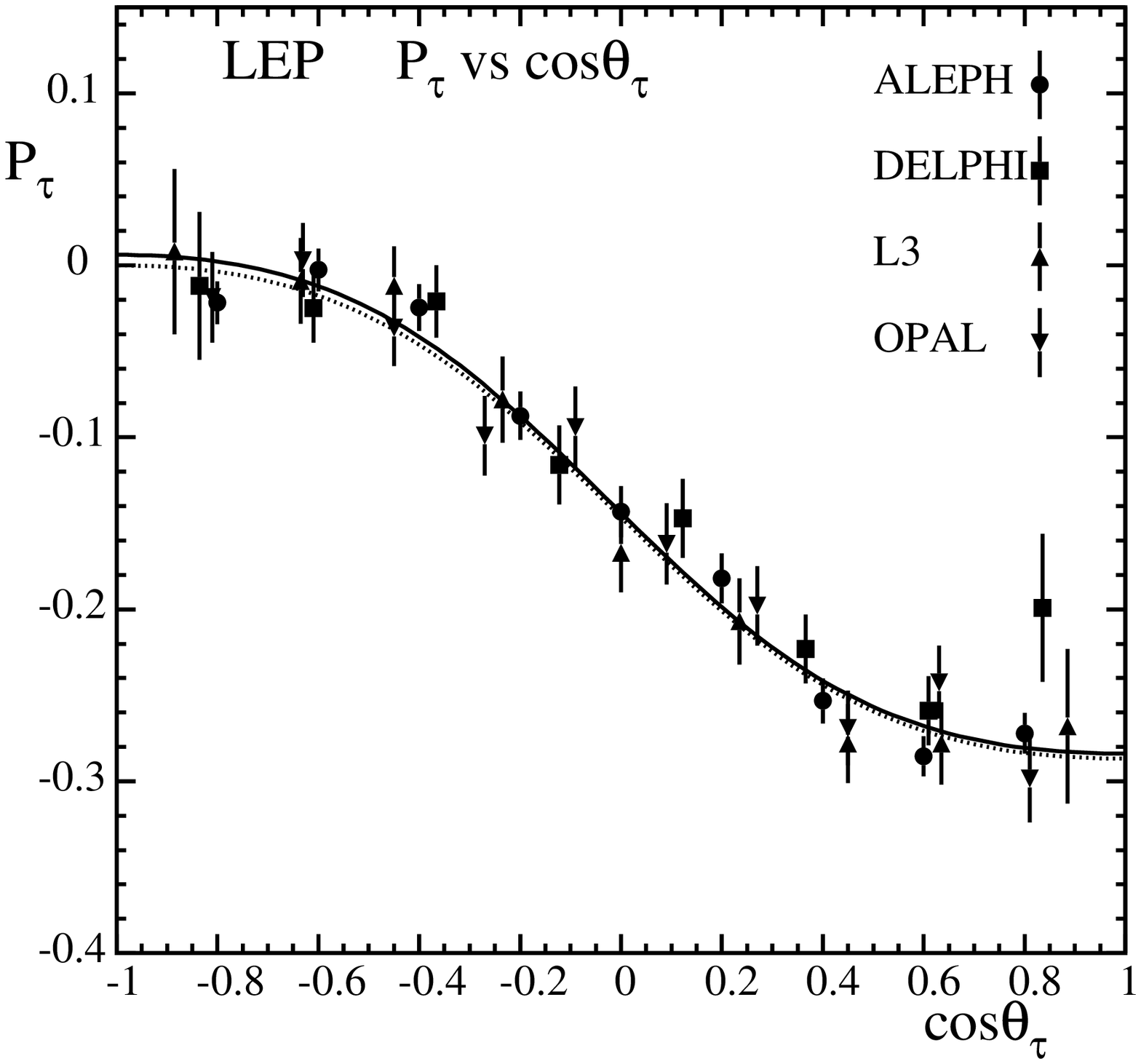}
\caption{Left: Contour curves of 68\% C.L. in the $(\RZl,\Afbzl)$
plane with and without lepton-universality. The SM expectation is
shown as the arrows for $\MT=174.3\pm5.1~\GeV$ and
$\MH=300^{+700}_{-186}~\GeV$, $\aqcd=0.118\pm0.002$, and, shifted for
visibility, $\dalhad=0.02761\pm0.0036$. Right: Tau polarisation as a
function of the polar scattering angle in tau-pair production at
LEP-1. The results of a fit to the data, with or without assuming
e-$\tau$ universality, are shown as the dashed and solid line.}
\label{fig:lsafb-ptau}
\end{center}
\end{figure}

The total cross section for hadron production as a function of the
$\Pee$ centre-of-mass energy in the vicinity of the Z pole is shown in
Figure~\ref{fig:ee-had} (right), comparing the measured cross sections
with those deconvoluted of QED effects which have to be known
precisely.  The $\chi^2$ per degrees of freedom are: 169/176 (ALEPH),
177/168 (DELPHI), 158/166 (L3), 155/194 (OPAL), and 36.5/31 for the
LEP combination, showing very good agreement when reducing the
hundreds of measurements to the few pseudo observables.  Assuming
lepton universality, the final results are~\cite{LEPLS}:
\begin{eqnarray}
\MZ ~ = ~ 91187.5\pm2.1~\MeV & \qquad & \GZ ~ = ~  2495.2\pm2.3~\MeV 
        \nonumber\\
\RZl~ = ~  20.767\pm0.025    & \qquad & \Afbzl ~ = ~ 0.0171\pm0.0010 \,.
\end{eqnarray}
The comparison of the measurements of $\RZl$ and $\Afbzl$ is shown in 
Figure~\ref{fig:lsafb-ptau} (left). 

The number of neutrinos, $\Nn=2.9841\pm0.0083$, is about 1.9 standard
deviations smaller than three. The uncertainty on $\Nn$ is given by:
\begin{eqnarray}
\delta\Nn & = & 10.5 \frac{\delta N_\Phad}{N_\Phad} ~\oplus~
                 3.0 \frac{\delta N_\Plep}{N_\Plep} ~\oplus~
                 7.5 \frac{\delta L      }{L      }\,,
\end{eqnarray}
where $\delta N_\Phad~(\delta N_\Plep)$ is the uncertainty on the
number of hadronic (leptonic) events, $\delta L$ is the uncertainty on
any absolute cross section measurement such as that arising from the
luminosity determination, and $\oplus$ denotes addition in
quadrature. Thus, the theoretical luminosity uncertainty of
0.06\%~\cite{StaszekJ} causes an uncertainty of $\pm0.0046$ on $\Nn$.

\subsection{Leptonic Polarisation Asymmetries}

In terms of the effective vector and axial-vector coupling constants,
$\gvf$ and $\gaf$, the asymmetry parameter $\Af$ is defined as:
\begin{eqnarray}
\Af & = & 2 \frac{\gvf/\gaf}{1+(\gvf/\gaf)^2} \,.
\end{eqnarray}
The leptonic asymmetry parameter is measured by SLD~\cite{MorrisS} and
at LEP-1 in various processes.  Assuming lepton universality, well
supported by the experimental results, the following final results are
obtained when combining the experiments:
\begin{eqnarray}
\Al & = & 0.1512\pm0.0042 \quad \hbox{forward/backward asymmetries}\\
\Al & = & 0.1465\pm0.0033 \quad \hbox{$\tau$ polarisation}\\
\Al & = & 0.1513\pm0.0021 \quad \hbox{left/right asymmetries (SLD)}\,,
\end{eqnarray}
showing good agreement, and with a combined value of:
\begin{eqnarray}
\Al & = & 0.1501\pm0.0016 \,.
\end{eqnarray}
The measurement of the $\tau$ polarisation as a function of polar
scattering angle is shown in Figure~\ref{fig:lsafb-ptau} (right),
where the LEP combination shows a $\chi^2$ per degree of freedom of
4.7/7.  For backward scattering, zero polarisation is expected and
observed.

\subsection{Heavy Flavour Results at the Z Pole}

\begin{figure}[h]
\begin{center}
\includegraphics[clip=true,bb=35 165 470 500,width=0.49\linewidth]{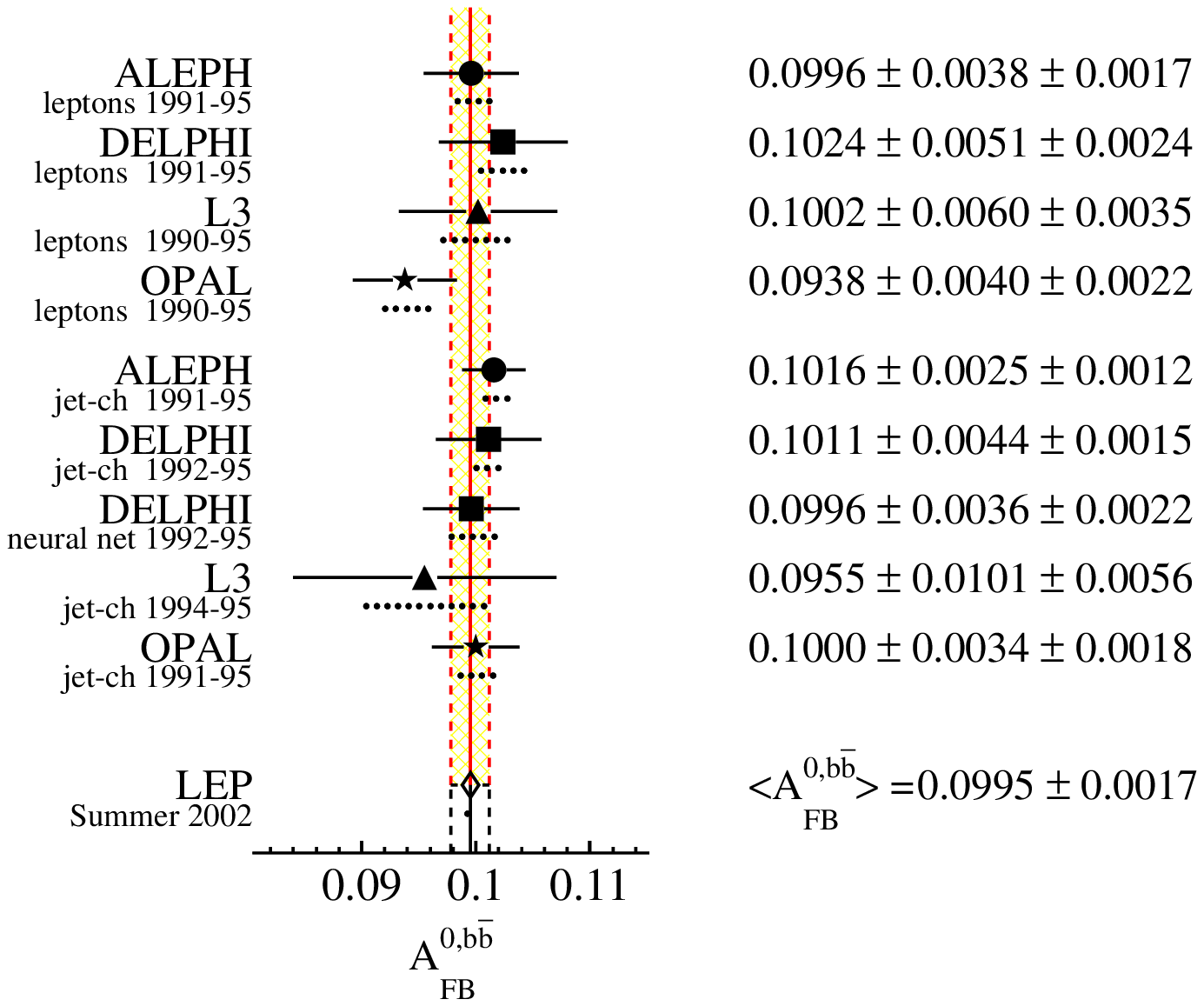}
\hfill
\includegraphics[clip=true,bb=35 165 470 500,width=0.49\linewidth]{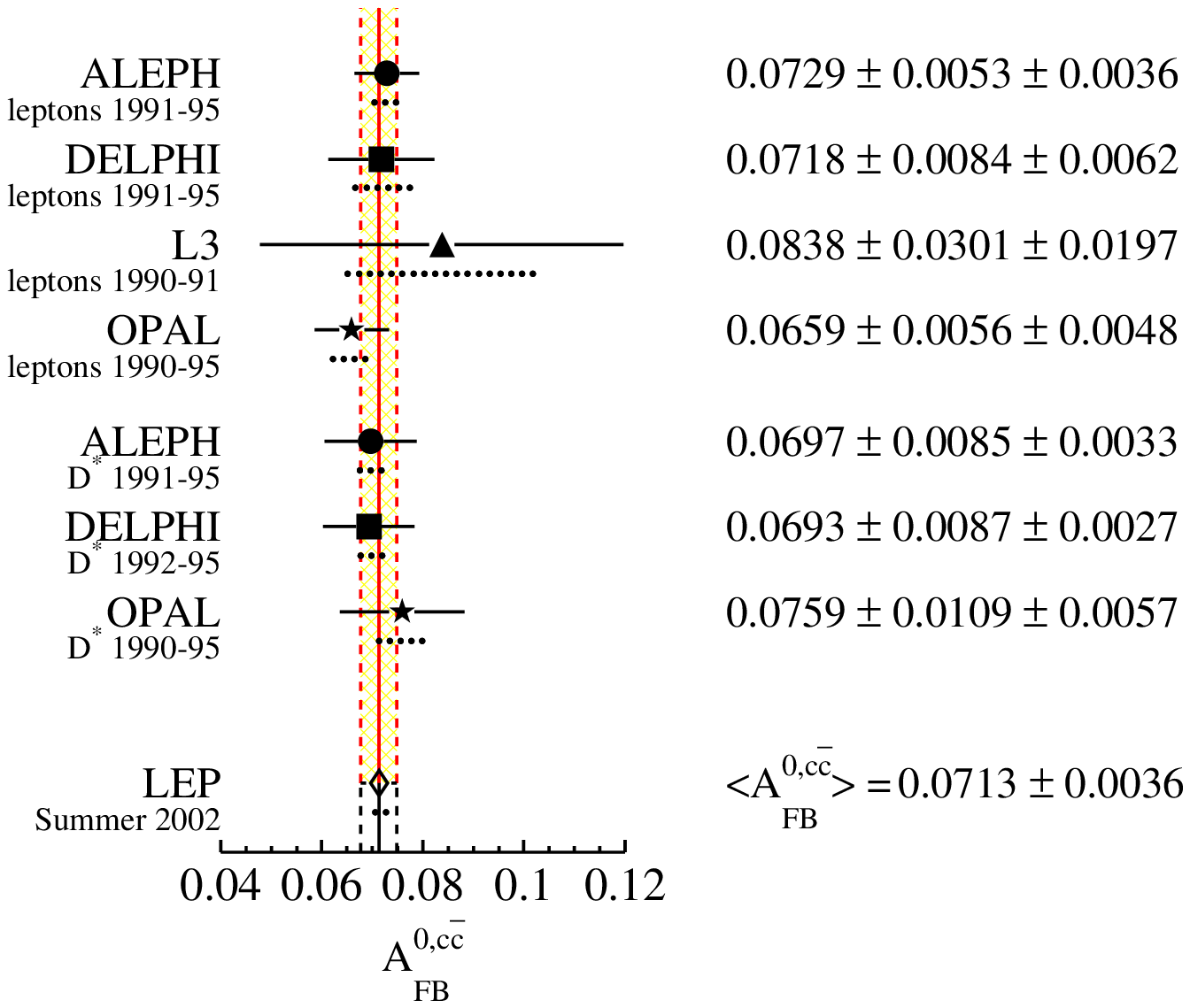}
\caption{Measurements of $\Afbzb$ (left) and $\Afbzc$ (right) at the Z pole.}
\label{fig:hf-bar-q}
\end{center}
\end{figure}

\begin{figure}[tbp]
\begin{center}
\includegraphics[width=0.49\linewidth]{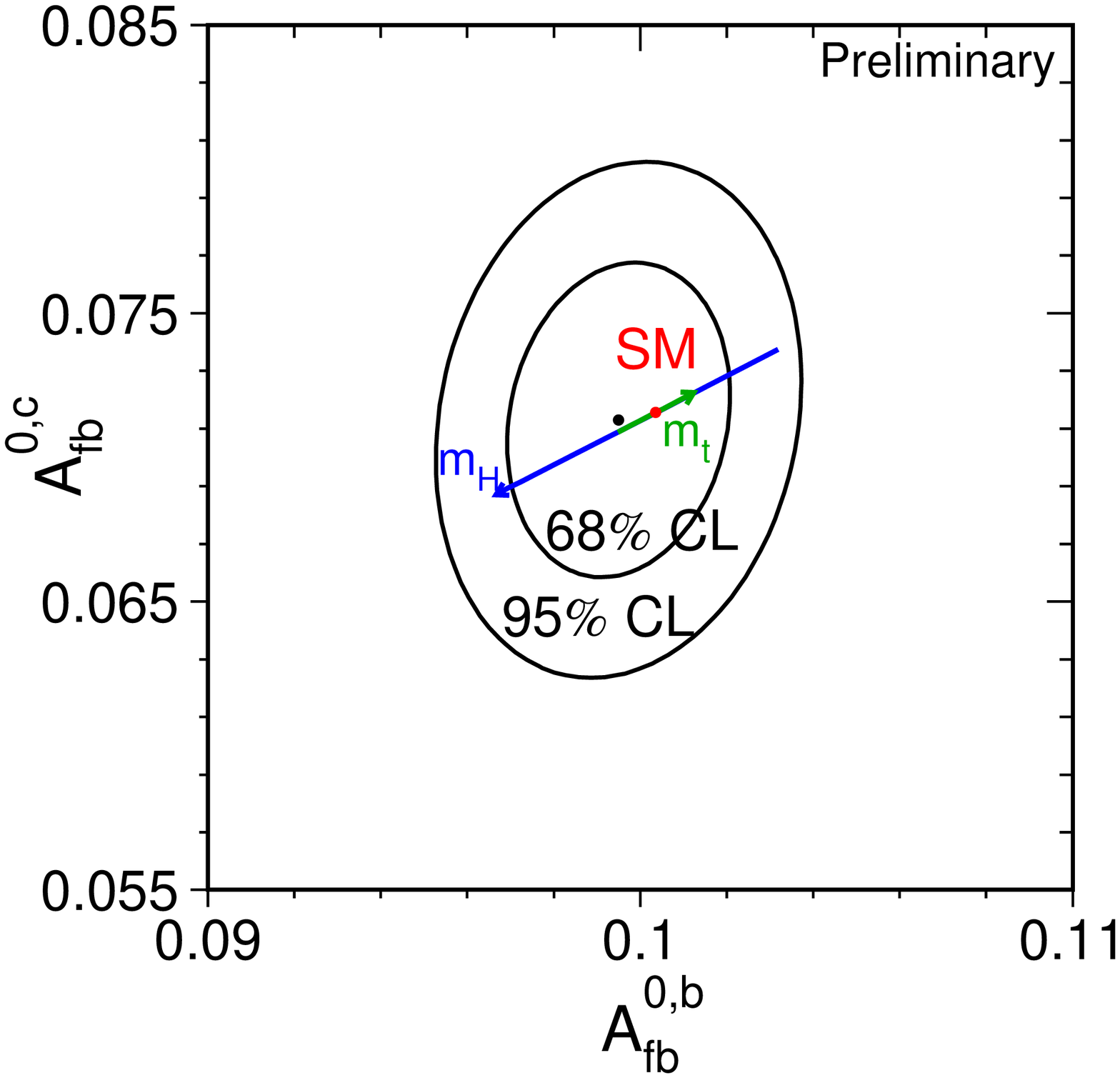}
\hfill
\includegraphics[width=0.49\linewidth]{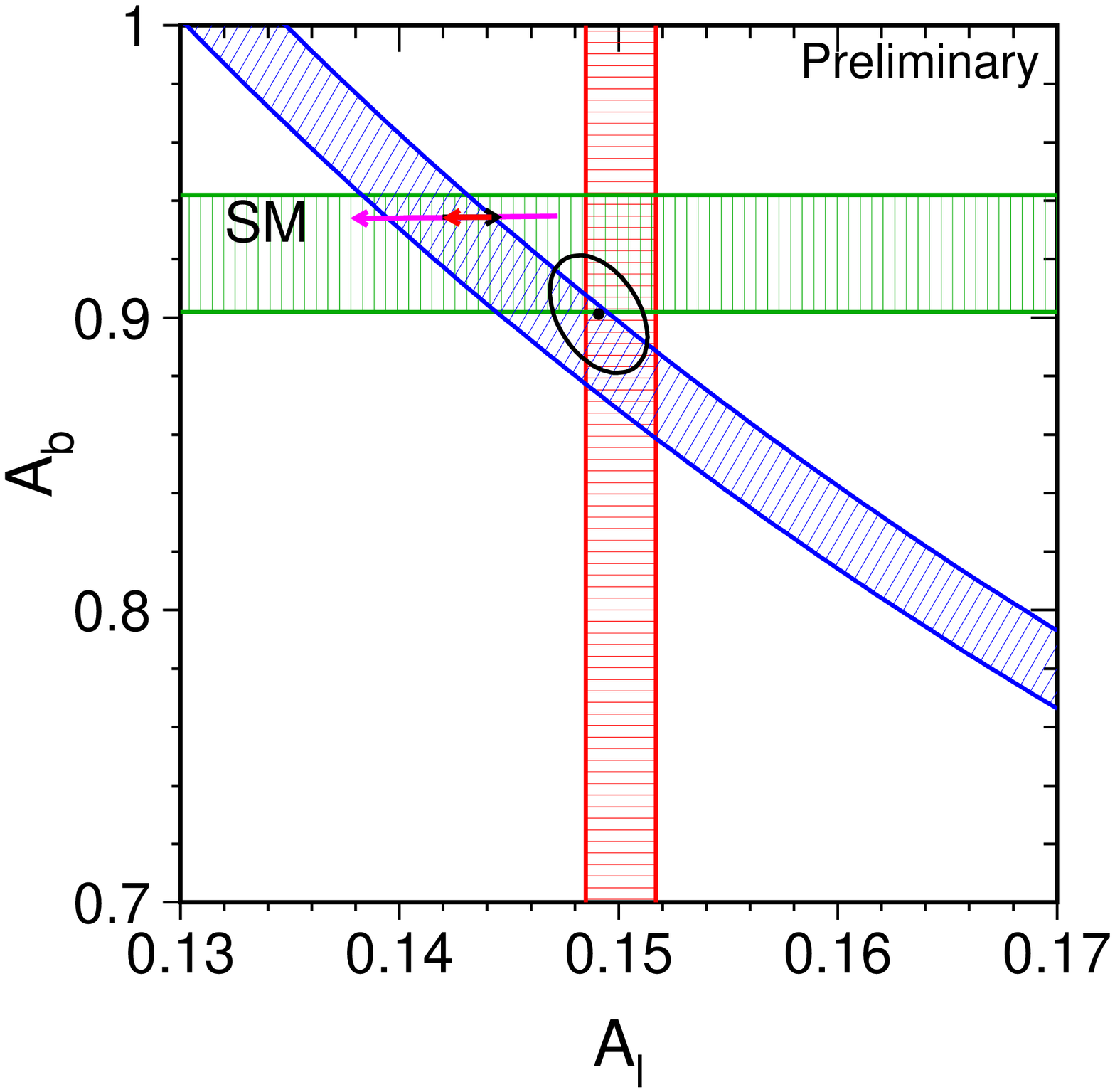}
\vskip -0.5cm
\caption{Left: Contour curve of 68\% C.L. in the $(\Afbzb,\Afbzc)$
  plane.  Right: Bands of $\pm1$ standard deviation width showing the
  combined results of $\Al$, $\Ab$, and $\Afbzb$.  The SM expectations
  are shown as the arrows for $\MT=174.3\pm5.1~\GeV$ and
  $\MH=300^{+700}_{-186}~\GeV$.}
\label{fig:coup:aq}
\end{center}
\end{figure}

While the results on b- and c-quark production rates
($\Rq=\GZq/\GZhad$) are final, several measurements of heavy-flavour
asymmetries by SLD and at LEP are still preliminary, and thus is the
joint combination of all heavy flavour results. Details on the various
heavy-flavour measurements at the Z pole are given
in~\cite{PippaW,WvanN}.  The combination has a rather low $\chi^2$ of
47.6 for $(105-14)$ degrees of freedom: all forward-backward
asymmetries are very consistent, as shown in
Figure~\ref{fig:hf-bar-q}, and their combination is still statistics
limited.  The combined values for $\Afbzb$ and $\Afbzc$ are compared
to the SM expectation in Figure~\ref{fig:coup:aq} (left), showing that
$\Afbzb$ agrees well with the SM expectation for an intermediate
Higgs-boson mass of a few hundred $\GeV$.

The mutual consistency of the measurements of $\Ab$,
$\Afbzb=(3/4)\Ae\Ab$ and $\Al$ assuming lepton universality is shown
in Figure~\ref{fig:coup:aq} (right).  Compared to the experimental
uncertainties, the SM predictions are nearly constant in $\Aq$, in
contrast to the situation for $\Al$.  This is a consequence of the SM
values of electric charge and iso-spin of quarks.  

\subsection{Effective Electroweak Mixing Angle}

Assuming the SM structure of the effective coupling constants, the
measurements of the various asymmetries are compared in terms of
$\swsqeffl$ in Figure~\ref{fig:sef2-mt-mw} (left).  The average of all
six $\swsqeffl$ determinations is:
\begin{eqnarray}
\swsqeffl & = & 0.23148\pm0.00017\,,
\label{eq:coup:swsqeffl}
\end{eqnarray}
with a $\chi^2/dof$ of 10.2/5, corresponding to a probability of
7.0\%. The enlarged $\chi^2/dof$ is solely driven by the two most
precise determinations of $\swsqeffl$, namely those derived from the
measurements of $\Al$ by SLD, dominated by the left-right asymmetry
result, and of $\Afbzb$ at LEP.  These two measurements differ by 2.9
standard deviations.  This is a consequence of the same effect as
shown in Figure~\ref{fig:coup:aq} (right).

\section{Global Standard Model Analysis}
\label{sec:MSM}

Within the framework of the SM, each pseudo observable presented above
is calculated as a function of five main relevant parameters, which
are the running electromagnetic and strong coupling constant evaluated
at the Z pole, $\alpha_{em}$ and $\aqcd$, and the masses of Z boson,
top quark and Higgs boson, $\MZ$, $\MT$, $\MH$.  Using the Fermi
constant $\GF$ allows to calculate the mass of the W boson. The
electromagnetic coupling is represented by the hadronic vacuum
polarisation $\dalhad$, as it is this contribution which has the
largest uncertainty.

\begin{figure}[tbp]
\begin{center}
\includegraphics[width=0.42\linewidth]{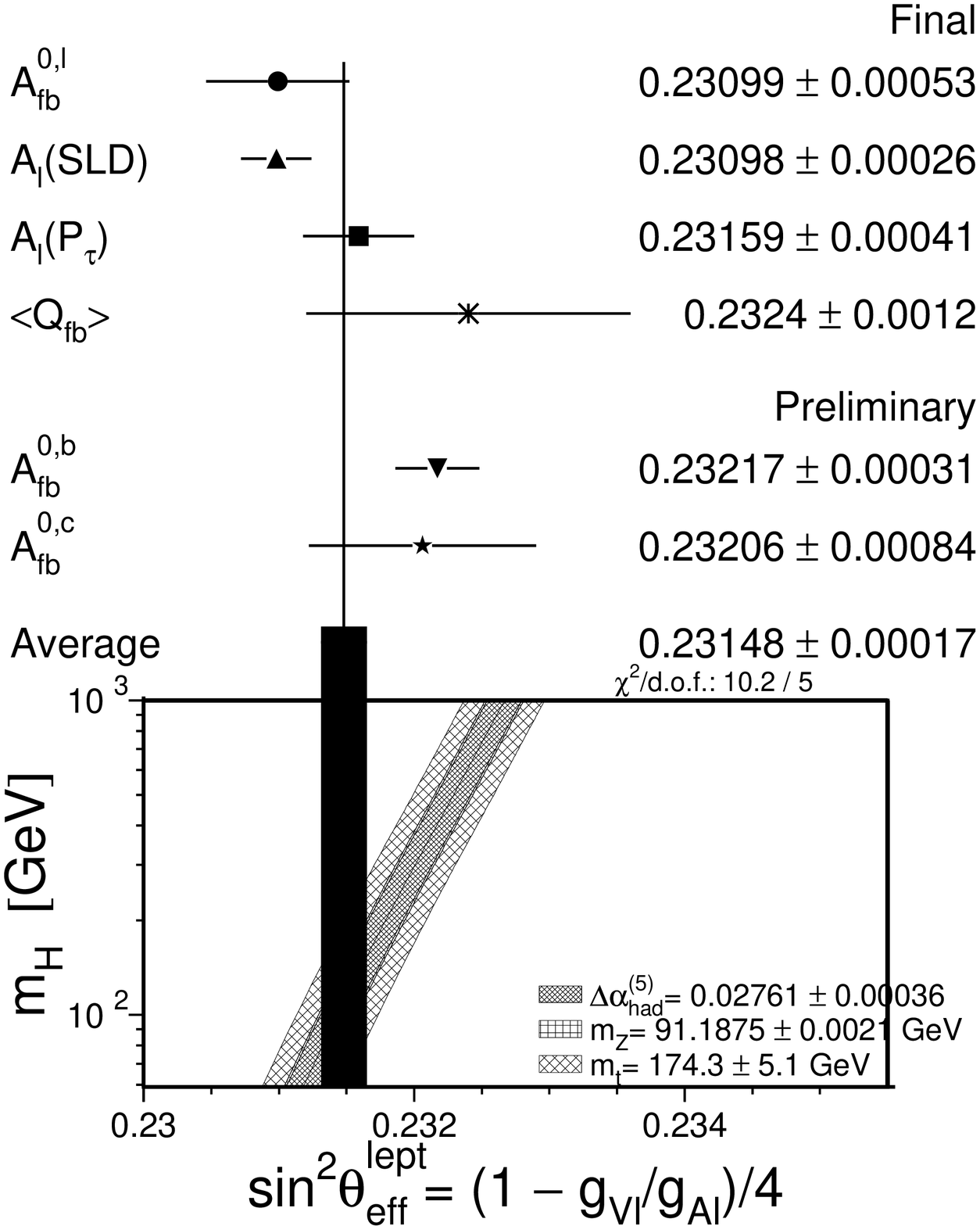}
\includegraphics[width=0.57\linewidth]{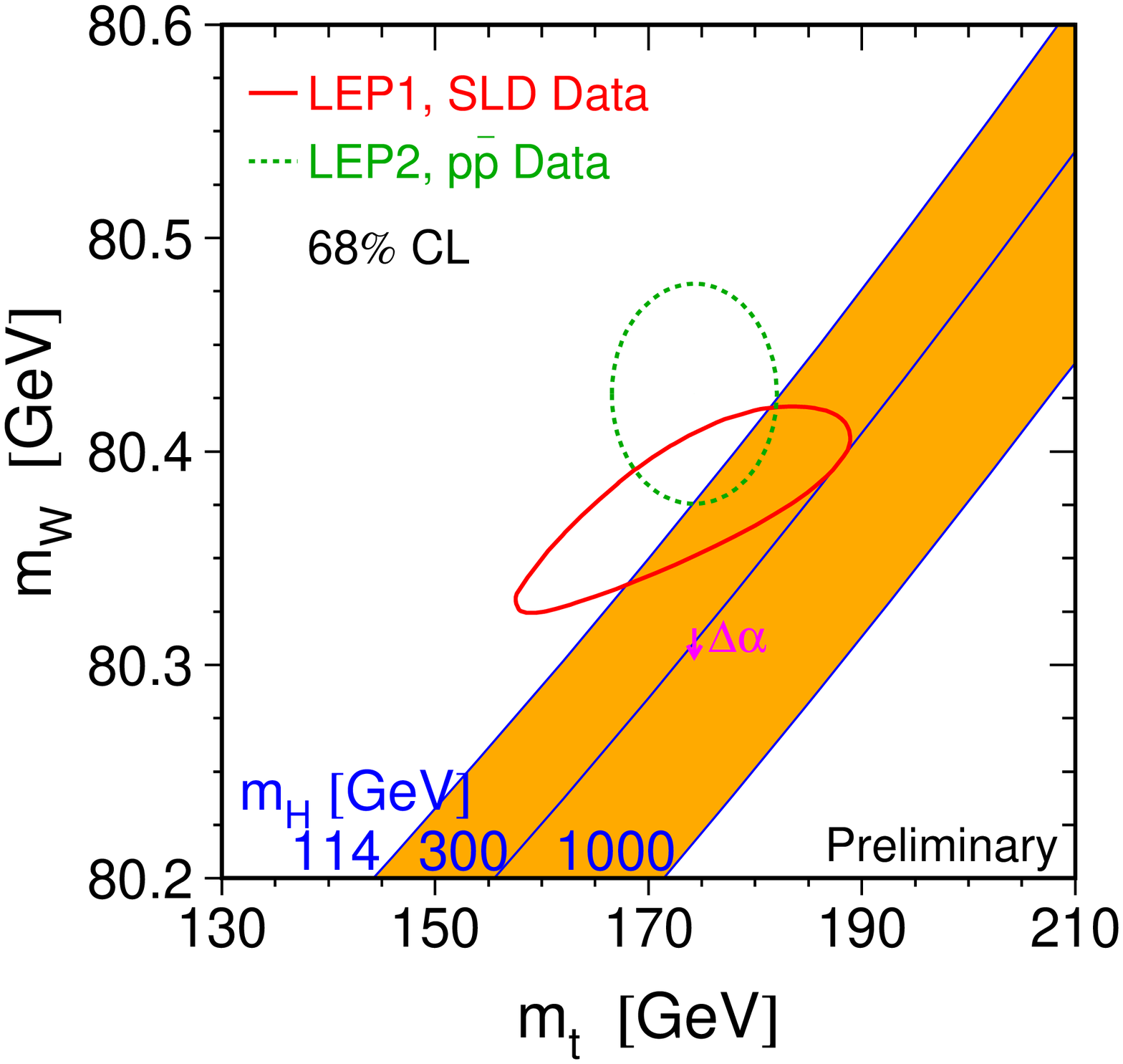}
\caption{Left: The effective electroweak mixing angle derived from
  various asymmetry measurements. Right: Contour curves of 68\%
  C.L. on the $(\MT,\MW)$ plane, for the corresponding direct
  and the indirect determinations. Also shown is the correlation
  between $\MW$ and $\MT$ as expected in the minimal SM for different
  Higgs boson masses. }
\label{fig:sef2-mt-mw}
\end{center}
\end{figure}

The precision of the Z-pole measurements require matching precision of
the theoretical calculations, first and second order electroweak and
QCD corrections etc. The dependence on $\MT$ and $\MH$ enters through
radiative corrections.  The predictions as a function of the five SM
parameters are calculated with the computer programs
TOPAZ0~\cite{TOPAZ0} and ZFITTER~\cite{ZFITTER}, which incorporate
state-of-the-art calculations, constraining the hadronic vacuum
polarisation to: $\dalhad=0.02761\pm0.00036$~\cite{BOLEK}.

Using the Z-pole measurements of SLD and LEP-1 in order to evaluate
electroweak radiative corrections, the masses of two heavy particles
measured at the TEVATRON and at LEP-2, namely the top quark and the W
boson, can be predicted. The resulting 68\% C.L. contour curve in the
$(\MT,\MW)$ plane is shown in Figure~\ref{fig:sef2-mt-mw}
(right). Also shown is the contour curve corresponding to the direct
measurements of both quantities at the TEVATRON and at LEP-2. The two
contour curves overlap, successfully testing the SM at the level of
electroweak radiative corrections. The diagonal band in
Figure~\ref{fig:sef2-mt-mw} (right) shows the constraint between the
two masses within the SM, which depends on the mass of the Higgs
boson, and to a small extent also on the hadronic vacuum polarisation
(small arrow labelled $\Delta\alpha$).  Both the direct and the
indirect contour curves prefer a low value for the mass of the SM
Higgs boson.

\begin{figure}[tbp]
\begin{center}
\includegraphics[width=0.42\linewidth]{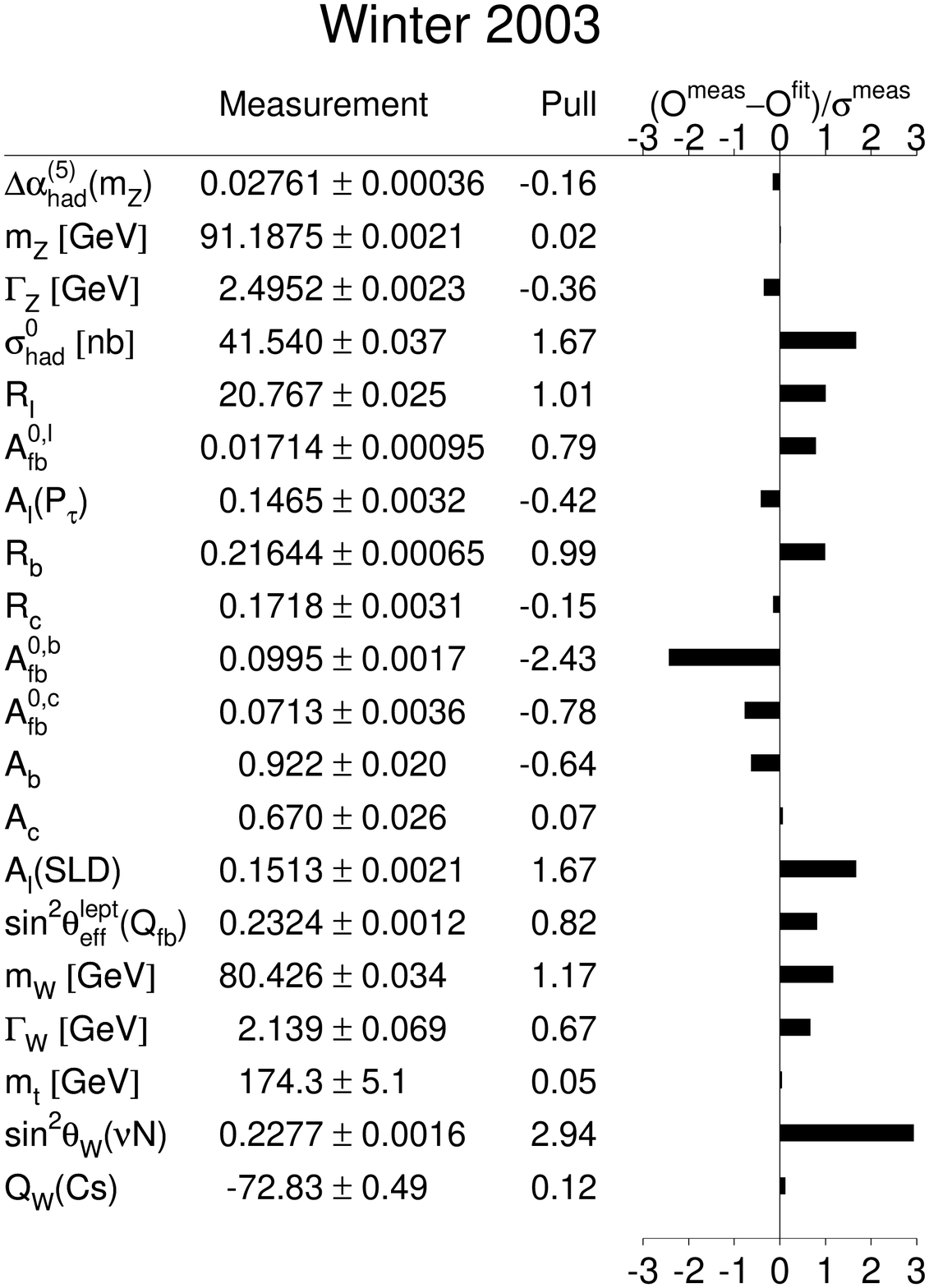}
\hfill
\includegraphics[width=0.57\linewidth]{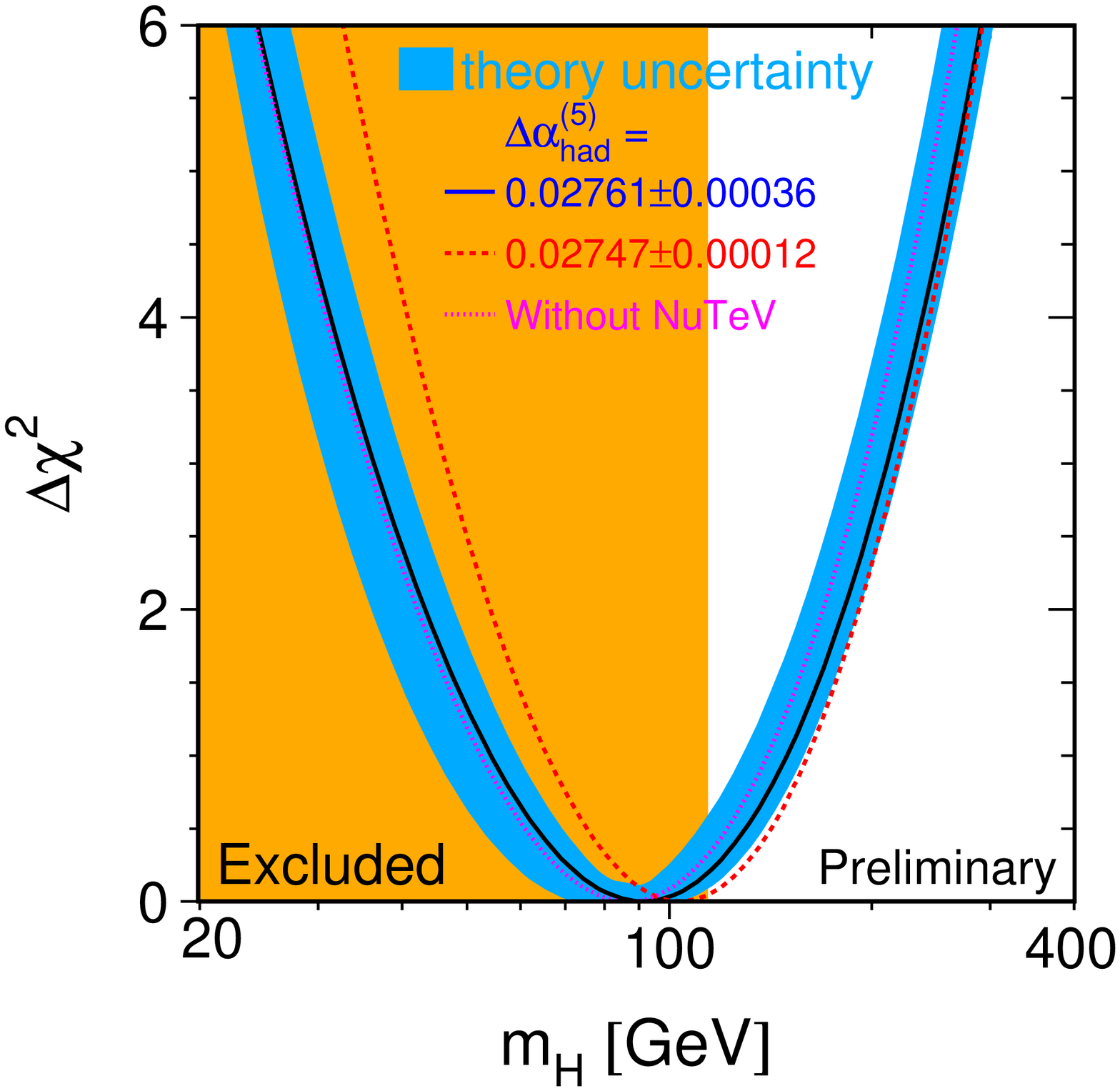}
\vskip -0.5cm
\caption{Left: Pulls of all 20 measurements used in the global SM
  analysis. The pull is the difference between measured and expected
  value calculated for the minimum of the $\chi^2$, divided by the
  measurement error. Right: $\Delta\chi^2$ curve as a function of
  $\MH$.  Also shown are the curves using a theory-driven evaluation
  of the hadronic vacuum polarisation, or excluding the NuTeV
  measurement.  }
\label{fig:pulls-blue}
\end{center}
\end{figure}

The best constraint on $\MH$ is obtained by analysing all data.  This
joint fit has a $\chi^2$ of 25.5 for 15 degrees of freedom,
corresponding to a probability of 4.4\%. The pulls of the 20
measurements entering the fit are shown in Figure~\ref{fig:pulls-blue}
(left).  The single largest contribution to the $\chi^2$, about 9
units, arises from the NuTeV measurement of the on-shell electroweak
mixing angle.  Excluding the NuTeV measurement, the $\chi^2/dof$
becomes 16.7/14, corresponding to 27.3\%, while the fitted parameters
in terms of central value and error are almost unchanged, showing that
the fit is robust against the NuTeV result. The second largest pull
arises from the $\Afbzb$ measurement as discussed above.

The global fit yields $\MH = 91^{+58}_{-37}~\GeV$, which corresponds
to a one-sided upper limit at 95\% C.L. on $\MH$ of $211~\GeV$
including the theory uncertainty as shown in
Figure~\ref{fig:pulls-blue} (left).  The fitted $\MH$ is strongly
correlated with the hadronic vacuum polarisation (correlation of
$-0.5$) and the fitted top-quark mass ($+0.7$). The strong correlation
with $\MT$ implies a shift of 35\% in the predicted $\MH$ if the
measurement of $\MT$ changes by one standard deviation
($5~\GeV$). Thus a precise experimental measurement of $\MT$ is very
important.

Also shown are the $\chi^2$ curves obtained with the theory-driven,
thus more precise evaluation of the hadronic vacuum
polarisation~\cite{YNDURAIN}, yielding also a reduced correlation of
$-0.2$ with $\MH$, or excluding the NuTeV result. Both analyses yield
nearly the same upper limits on $\MH$.  In case the measurement with
the second largest pull, $\Afbzb$, is removed, the 95\% C.L. upper
limit on $\MH$ reduces to $149~\GeV$ including the theory uncertainty.
The theoretical uncertainty on the SM calculations of the observables
is visualised as the thickness of the blue band. It is dominated by
the theoretical uncertainty in the calculation of the effective
electroweak mixing angle, where a two-loop calculation is needed.

The shaded part in Figure~\ref{fig:pulls-blue} (left) shows the $\MH$
range up to $114.4~\GeV$ excluded by the direct search for the Higgs
boson at 95\% confidence level. Even though the minimum of the
$\chi^2$ curve lies in the excluded region, the uncertainties on the
Higgs mass value are such as that the results are well compatible.
Further discussions are presented in~\cite{MikeCh}.

\section{Caveats: Low Higgs-Boson masses}

\begin{figure}[ht]
\begin{center}
\includegraphics[width=0.42\linewidth]{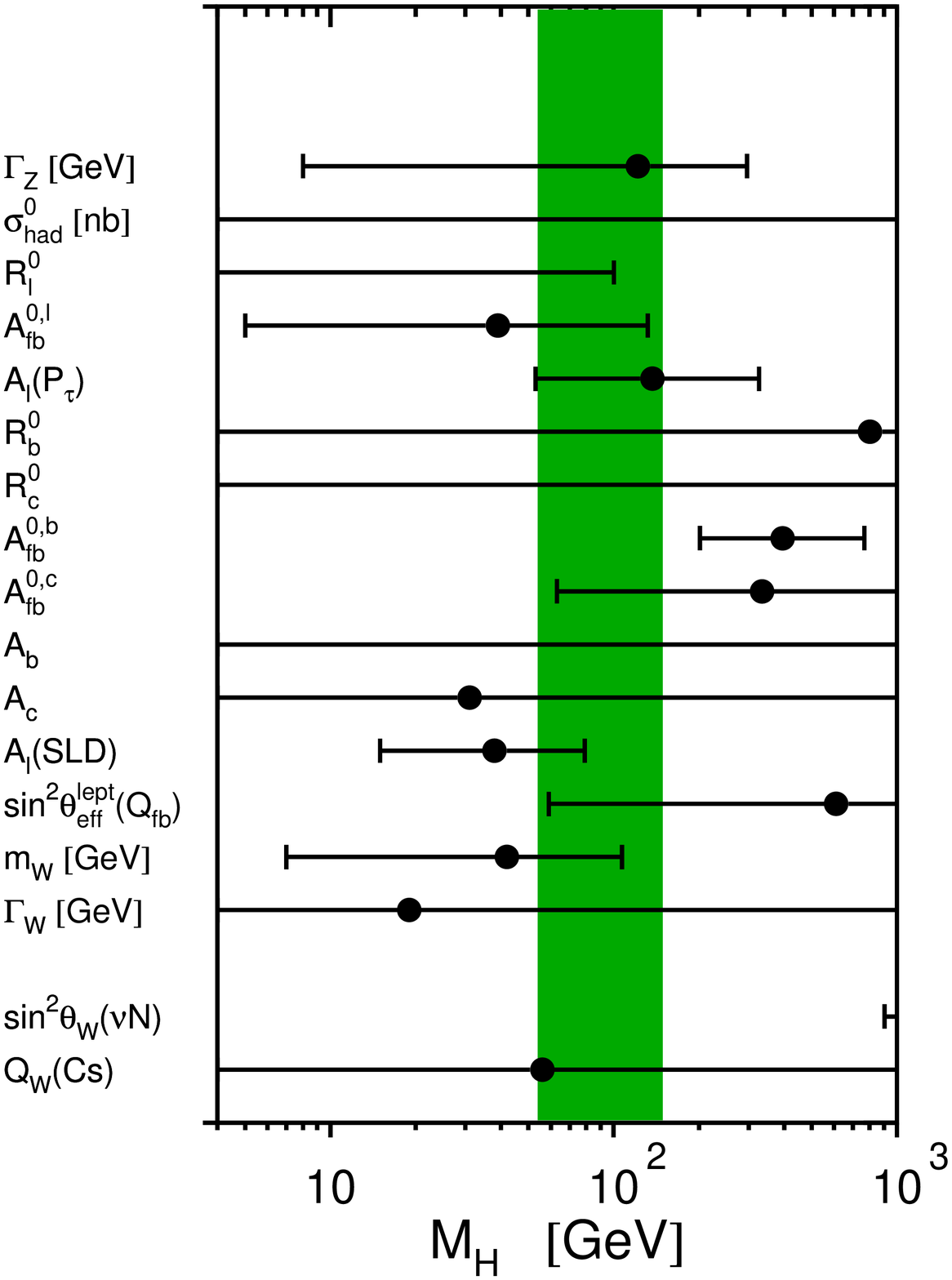}
\hfill
\includegraphics[width=0.57\linewidth]{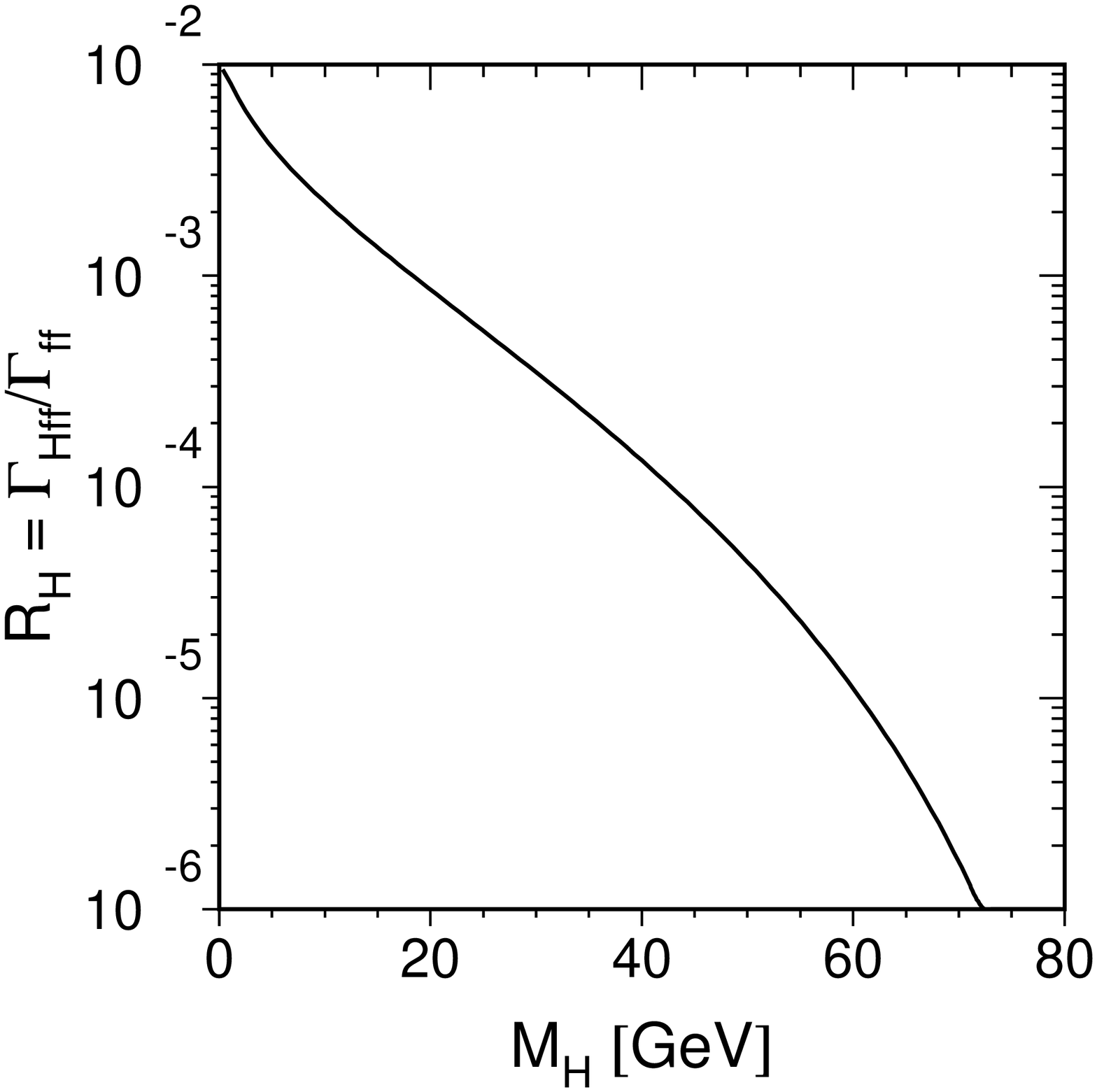}
\vskip -0.5cm
\caption{Left: Higgs mass constraint from each pseudo observable.
  Right: ratio $\RH = \Gamma(\PZ^*\PH)/\GZ$ as a function of the
  Higgs-boson mass. }
\label{fig:higgs}
\end{center}
\end{figure}

The measurement of a pseudo observable may also be interpreted as a
constraint on the mass of the Higgs boson.  In order to evaluate this
constraint, a full five-parameter Standard-Model fit is performed,
constraining the four SM input parameter besides $\MH$ as follows:
$\dalhad=0.02761\pm0.00036$, $\aqcd=0.118\pm0.002$,
$\MZ=91187.5\pm2.1~\MeV$ and $\MT=174.3\pm5.1~\GeV$.  The resulting
$\MH$ constraints are shown in Figure~\ref{fig:higgs} (left).

However, care must be taken when interpreting results quantitatively in
the region of low values of the Higgs mass. Neither the experimental
analyses extracting the pseudo observables from the raw data, nor the
SM calculation of the predictions take into account the real
Higgs-strahlungs process $\Pee\to\PZ\to\PZ^*\PH\to\Pff\PH$.  This
effect can be quite sizeable: the fraction of real Higgs-strahlung,
$\RH=\Gamma(\PZ^*\PH)/\GZ$, is shown in Figure~\ref{fig:higgs}
(right), ranging up to 1\% for very low Higgs
masses~\cite{LEP1YRVOL2}.  

If all $\PZ^*\PH$ decay modes were selected as hadronic events, then
the measured total and hadronic width would increase by the amount
$\RH\GZ$. This increase would lead to an increase of
$\Delta\aqcd\simeq4\RH$ in the fitted value in analyses neglecting to
consider $\PZ^*\PH$.  The requirement that the shift in $\aqcd$ should
be less than a 10\% of the fitted error on $\aqcd$ forces
$\MH>22~\GeV$. Since also the other fit parameters are unaffected,
this does not pose a problem.  The specific heavy-flavour analyses may
also be affected. The maximal effect is observed if all $\PZ^*\PH$
decay modes are tagged a b-quark events. Then $\GZb$ is increased by
the same amount as $\GZhad$, and $\Afbzb$ is changed. This is most
visible in a shift of $\RZb$: $\Delta\RZb \simeq 1.1\RH$. The
requirement that this shift is less than 10\% of the experimental
error on $\Rb$ forces $\MH>47~\GeV$.  The reconstruction of $\MW$ at
LEP-2 is probably not affected. However, in addition the effect of
$\PZ^*\PH$ is centre-of-mass energy dependent.  While the global SM
fit is valid for the central value of $\MH$ and upper errors,
quantitative statements in the low-Higgs mass regime remain dubious. A
fully correct treatment would require experimental efficiencies and
corrections for $\PZ^*\PH$ as a function of $\MH$ which are not
available. In any case, the limit from the direct search for the Higgs
boson of $\MH>114.4~\GeV$ at 95\% confidence level~\cite{SMHiggs} is
sufficiently high.

\section{Conclusions}

During the last 15 years many experiments at both hadron and lepton
collider have performed a wealth of measurements with unprecedented
precision in high-energy particle physics. These measurements test all
aspects of the SM of particle physics, and many of them show large
sensitivity to electroweak radiative corrections at loop level.

Most measurements agree with the expectations as calculated within the
framework of the SM, successfully testing the SM at Born and at loop
level. Still there are two ``2.9 standard deviations effects'', namely
the spread in the various determinations of the effective electroweak
mixing angle, within the SM analysis apparently disfavouring the
measurement of $\Afbzb$, and NuTeV's $R_-$ result, most pronounced
when interpreted in terms of the on-shell electroweak mixing angle.

The validity of any pseudo-observable analysis rests on the assumption
that the effects of real Higgs production (or that of any non-SM final
state) must be negligible.

For the future, precise theoretical calculations including theoretical
uncertainties~\cite{WvanN,TriTH} are needed, in particular a two-loop
calculation for the effective electroweak mixing
angle. Experimentally, the next few years will bring improvements in
the measurements of W and top masses, and the long-awaited discovery
of the Higgs boson.

\section*{Acknowledgements}

It is a pleasure to thank my colleagues of the TEVATRON and LEP
electroweak working groups, members of the NuTeV, SLD, ALEPH, DELPHI,
L3, OPAL, CDF and D\O\ experiments, as well as D.~Bardin, G.~Passarino
and G.~Weiglein for valuable discussions.


\begin{thebibliography}{99}

\bibitem{LEPEWWG}     The LEP Electro-Weak Working Group (EWWG), 
                      hep-ex/0212036 and winter 2002/03 update LEPEWWG/2003-01,
                      {\tt http://www.cern.ch/LEPEWWG}.
\bibitem{FredJ}       F.~Jegerlehner, {\em these proceedings}.
\bibitem{APV-Caesium} C.S.~Wood \etal, Science 275 (1997) 1759; 
                      M.Y.~Kuchiev, J.Phys. B35 (2002) 503.
\bibitem{PWR}         E.~Paschos, L. Wolfenstein, PRD 7 (1973) 91.
\bibitem{NuTeV}       G.P.~Zeller \etal, PRL 88 (2002) 091802.
\bibitem{nNWorld} K.S.~McFarland, proceedings 28th ICHEP, Warsaw,
                      Poland (1996); M. Goncharov \etal, PRD 64 (2001) 112006.
\bibitem{KevinMcF}    K.S.~McFarland, {\em these proceedings}.
\bibitem{SMoch}       S.~Moch, {\em these proceedings}.
\bibitem{RPP2002}     Review of Particle Properties 2002, PRD, Vol.66, No.1., 
                      010001.
\bibitem{D0-MT-L+J} J.~Estrada, talk on behalf of the D\O\
collaboration, HCP02 - 14th Topical Conference On Hadron Collider
Physics, Karlsruhe, Germany, September 2002, hep-ex/0302031.

\bibitem{TEV-MW-GW} 
Tevatron Electroweak Working Group and the CDF
and D\O\ Collaborations, CDF Note 5888, D\O\ Note 3963,
FERMILAB-FN-0716, July 2002.

\bibitem{UlrichB}     U.~Baur, {\em these proceedings}.
\bibitem{RichardH}    R.~Hawkings, {\em these proceedings}.

\bibitem{LEPLS}       The LEP Collaborations and the 
                      LEP EWWG, hep-ex/0101027.

\bibitem{StaszekJ}    S.~Jadach, {\em these proceedings}.
\bibitem{MorrisS}     M.~Swartz,  {\em these proceedings}.
\bibitem{PippaW}      P.~Wells,    {\em these proceedings}.
\bibitem{WvanN}       W.~van Neerven, {\em these proceedings}.

\bibitem{TOPAZ0}      G.~Passarino \etal, CPC 117 (1999) 278.
\bibitem{ZFITTER}     D.~Bardin    \etal, CPC 133 (2001) 229.
\bibitem{BOLEK}       H.~Burkhardt, B. Pietrzyk, 
                                         PLB 513 (2001) 46.
\bibitem{YNDURAIN}    J.F.~de Troconiz, F.J.~Yndurain, 
                                         PRD 65 (2002) 093002
\bibitem{MikeCh}      M.~Chanowitz, {\em these proceedings}.
\bibitem{LEP1YRVOL2}  Z Physics at LEP 1, CERN yellow report
                      89-08, Vol.2, p.6.

\bibitem{SMHiggs}     The LEP Collaborations and the LEP HWG, 
                      CERN-EP/2003-011, subm.~to PLB.

\bibitem{TriTH}       W.~Hollik, {\em these proceedings}; 
                      G.~Passarino, {\em these proceedings};
                      G.~Weiglein, {\em these proceedings}.


\end{thebibliography}
\end{document}